\documentclass[twocolumn]{aastex701}

\usepackage{mathtools, amsmath}
\usepackage{nccmath}
\usepackage{float}
\usepackage{graphicx}
\usepackage{fontawesome}
\usepackage{hyperref}
\usepackage{placeins}
\usepackage{ulem}
\usepackage{multirow}
\usepackage{hhline}
\usepackage{xcolor}
\usepackage{tabularx,booktabs}
\hypersetup{
    colorlinks,
    linkcolor={red!50!black},
    citecolor={blue!50!black},
    urlcolor={blue!80!black}
}

\newcommand{\git}[2]{\href{https://github.com/#1}{\faGithub}\footnote{\label{#1}#2\url{https://github.com/#1}.}}

\newcommand{\fsig}{f\sigma_8}
\newcommand{\sn}{SN~Ia}
\newcommand{\sns}{SNe~Ia}

\newcommand{\simu}[1]{$S_\mathrm{#1}$}

\newcommand{\DUKE}{\affiliation{Department of Physics, Duke University, Durham, NC 27708, USA}}
\newcommand{\CPPM}{\affiliation{Aix Marseille Univ, CNRS/IN2P3, CPPM, Marseille, France}}

\defcitealias{carreres_growth-rate_2023}{C23}
\defcitealias{vincenzi_dark_2024}{V24}
\defcitealias{popovic_improved_2021}{P21}
\defcitealias{brout_its_2021}{BS21}
\defcitealias{popovic_pantheon_2023}{P23}

\begin{document}

\title{Type Ia supernova growth-rate measurement with LSST simulations: intrinsic scatter systematics}

\author[0000-0002-7234-844X]{Bastien Carreres}
\DUKE
\email[show]{bastien.carreres@duke.edu}

\author[0000-0003-3917-0966]{Rebecca C. Chen}
\DUKE
\email{}

\author[0000-0001-8596-4746]{Erik R. Peterson}
\DUKE
\email{}

\author[0000-0002-4934-5849]{Dan Scolnic}
\DUKE
\email{}

\author[0000-0002-3500-6635]{Corentin Ravoux}
\affiliation{Université Clermont-Auvergne, CNRS, LPCA, 63000 Clermont-Ferrand, France}
\email{}

\author[0000-0001-6839-1421]{Damiano Rosselli}
\CPPM
\email{}

\author[0000-0002-5389-7961]{Maria Acevedo}
\DUKE
\email{}

\author[0000-0002-9885-3989]{Julian E. Bautista}
\CPPM
\email{}

\author[0000-0002-7496-3796]{Dominique Fouchez}
\CPPM
\email{}

\author[0000-0002-1296-6887]{Lluís Galbany}
\affiliation{Institute of Space Sciences (ICE-CSIC), Campus UAB, Carrer de Can Magrans, s/n, E-08193 Barcelona, Spain}
\affiliation{Institut d'Estudis Espacials de Catalunya (IEEC), 08860 Castelldefels (Barcelona), Spain}
\email{}

\author[0000-0001-8861-3052]{Benjamin Racine}
\CPPM
\email{}

\author[]{The LSST Dark Energy Science Collaboration}
\affiliation{SLAC National Accelerator Laboratory, Menlo Park, CA 94025, USA}
\email{}

\correspondingauthor{Bastien Carreres}

\begin{abstract}
Measurement of the growth rate of structures ($\fsig$) with Type Ia supernovae (\sns) will improve our understanding of the nature of dark energy and enable tests of general relativity. In this paper, we generate simulations of the 10 year \sn\ dataset of the Rubin-LSST survey, including a correlated velocity field from a N-body simulation and realistic models of \sns\ properties and their correlations with host-galaxy properties.
We find, similar to SN~Ia analyses that constrain the dark energy equation-of-state parameters $w_0w_a$, that constraints on $\fsig$ can be biased depending on the intrinsic scatter of \sns. While for the majority of intrinsic scatter models we recover $\fsig$ with a precision of $\sim$13--14\%, for the most realistic dust-based model, we find that the presence of non-Gaussianities in Hubble diagram residuals leads to a bias on $\fsig$ of about $\sim-20\%$. When trying to correct for the dust-based intrinsic scatter, we find that the propagation of the uncertainty on the model parameters does not significantly increase the error on $\fsig$. We also find that while the main component of the error budget of $\fsig$ is the statistical uncertainty ($>75\%$ of the total error budget), the systematic error budget is dominated by the uncertainty on the damping parameter, $\sigma_u$, that gives an empirical description of the effect of redshift space distortions on the velocity power spectrum. Our results motivate a search for new methods to correct for the non-Gaussian distribution of the Hubble diagram residuals, as well as an improved modeling of the damping parameter.
\end{abstract}


\section{Introduction}
In the past decades, Type Ia supernovae (\sns) have been used as powerful tools to probe the content of our universe and test the standard model of cosmology. In particular, they have been widely used in measurements of the Hubble constant, $H_0$ (e.g. \citealt{freedman_carnegie-chicago_2019,riess_comprehensive_2022, scolnic_cats_2023, galbany_updated_2023}), and of the dark energy equation-of-state parameter, $w$ \citep{brout_pantheon_2022, des_collaboration_dark_2024}. The number of available \sns\ for cosmology have grown from a few hundred in the 2010's to a few thousand today with datasets consisting of multiple surveys \citep{scolnic_pantheon_2022, sanchez_dark_2024}. This number is expected to increase by several orders of magnitude, reaching $\sim O(100~000)$ \sns\ in the coming years with a new generation of full-sky surveys such as the Zwicky Transient Facility (ZTF, \citealt{dhawan_zwicky_2021, rigault_ztf_2025-1}) in the Northern Hemisphere and the Vera C. Rubin Observatory Legacy Survey of Space and Time\footnote{\url{http://www.lsst.org}.} (LSST) in the Southern Hemisphere. These datasets will enable the use of \sns\ in constraining new cosmological parameters such as the product of the growth rate of structure $f$ with the amplitude of matter density fluctuations $\sigma_8$, commonly noted $\fsig$.
The growth rate of structure, $\fsig$, quantifies the rate of evolution of matter overdensities and velocity perturbations. Hence, this parameter is sensitive not only to dark energy, but also to the law of gravity and can be used to test the nature of dark energy and general relativity \citep{huterer_growth_2023, turner_cosmology_2024}. The measurement of the growth rate of structure is part of the scientific goals of the LSST Dark Energy Science Collaboration\footnote{\url{http://lsstdesc.org}.} (DESC, \citealt{lsst_dark_energy_science_collaboration_large_2012}). In this work, we prepare for the future measurement of $\fsig$ using LSST \sn\ data by investigating some of the key systematics that could impact the analysis.

The measurement of the growth rate of structure involves peculiar velocities (PVs). Peculiar velocities refer to galaxies motions of galaxies external to the Hubble flow velocities emerging from the Universe's expansion. They are imprinted in galaxy redshift measurements as an additional shift  of their spectrum wavelengths due to the relativistic Doppler effect. Thus, an observed redshift, $z_\mathrm{obs}$, can be decomposed into a combination of the cosmological expansion redshift, $z_\mathrm{cos}$, and the Doppler shift caused by PVs, $z_p$ as
\begin{equation}\label{eq:zobs}
    1 + z_\mathrm{obs} = (1 + z_\mathrm{cos})(1 + z_p),
\end{equation}
where the peculiar redshift $z_p$ can be approximated by $z_p\simeq v/c$ with $v$ the velocity of the galaxy projected on the line-of-sight and $c$ the velocity of light.

Other methods beyond SNe Ia have also been used to constrain $\fsig$ and can be divided into two main categories: indirect and direct measurements. Indirect measurements constrain $\fsig$ through the redshift space distortion (RSD) caused by the Doppler effect from PVs on the correlation function of galaxy clustering \citep{kaiser_clustering_1987}. The most recent measurements using this technique are those from the Dark Energy Spectroscopic Instrument (DESI) second data release \citep{desi_collaboration_desi_2025}, which is a spectroscopic survey mapping large-scale structures over a 14~000 deg$^2$ area of the sky.
These indirect measurements require a high number of galaxies and are more precise at high redshifts where there is more volume. Measurements of $\fsig$ through RSD of galaxy clustering are expected to give a percent-level precision at the end of the DESI survey \citep{desi_collaboration_desi_2016}.

On the other hand, direct measurements of $\fsig$ use PV estimations and apply summary statistics to them to extract cosmological constraints. This type of measurement requires, in addition to spectroscopic redshifts, estimation of distances from a photometric survey in order to break the degeneracy between $z_\mathrm{cos}$ and $z_p$ in Eq.~\ref{eq:zobs} and infer the galaxy PVs. This direct method is most efficient at low-$z$ where the amplitude of $z_p$ is non-negligible compared to the cosmological redshift $z_\mathrm{cos}$. Most of the existing direct constraints for $\fsig$ use galaxies as distance indicators through the Tully-Fisher (TF, \citealt{tully_new_1977}) and Fundamental Plane (FP, \citealt{djorgovski_fundamental_1987}) relationships. After the estimation of PVs, several summary statistics can be applied to constrain $\fsig$. These summary statistics include the compressed two-point correlation function  \citep{nusser_velocitydensity_2017,qin_redshift_2019,turner_local_2022}; the comparison between estimated velocities and velocities obtained from density field reconstruction \citep{carrick_cosmological_2015,boruah_cosmic_2020, said_joint_2020}; the forward modelling reconstruction of the velocity field \citep{boruah_reconstructing_2021, valade_hamiltonian_2022,prideaux-ghee_field-based_2022}; and the maximum likelihood method \citep{abate_peculiar_2009, johnson_6df_2014, howlett_2mtf_2017}. More recently, this has also been extended to cross-correlation analyses with PVs and density clustering data \citep{adams_joint_2020, lai_using_2023}.

In this work, we focus on the direct constraint of $\fsig$ with the maximum likelihood method using \sns\ as distance indicators. Type Ia supernovae are better distance indicators than galaxies, with a relative precision of about $\sim$~6\% compared to $\sim20\%$ for the distances inferred with TF and FP relationships. However, currently available low-$z$ \sn\ datasets are too small and not homogeneous enough to allow a \sns\ only measurement of $\fsig$, and most \sn\ cosmological analyses have instead treated PVs as a systematic \citep{davis_effect_2011,peterson_pantheon_2022, carr_pantheon_2022, carreres_ztf_2025,peterson_improving_2025,hollinger_uncertainties_2025, tsaprazi_field-level_2025}. A few analyses have made use of \sn\ data in combination with other probes to constrain $\fsig$ \citep{carrick_cosmological_2015,huterer_testing_2017,boruah_reconstructing_2021}, but their methods were not extensively tested with simulations. In the context of the ZTF and LSST surveys, multiple Fisher forecast studies have been done and predict a precision for $\fsig$ of a few percent with \sns\ only and in combination with density obtained from spectroscopic surveys \citep{koda_are_2014,howlett_cosmological_2017,graziani_peculiar_2020,kim_complementarity_2020}. The first investigation of a \sns\ only measurement of $\fsig$ with simulations is presented in \cite{carreres_growth-rate_2023} (hereafter C23) within the context of the ZTF survey. This analysis showed that the 6 years of the ZTF survey can provide a constraint on $\fsig$ with an error of 20\% using a spectroscopically classified and complete ($z<0.06$) \sns\ sample. However, \citetalias{carreres_growth-rate_2023} mainly focused on the selection function bias systematics, and additional systematics were neglected and left for future work. For instance, the simulations used in \citetalias{carreres_growth-rate_2023} are based on a dark matter halo $N$-body simulation and do not attempt to model correlations between \sns\ and their host galaxies. Furthermore, the contamination of the \sn\ sample by core-collapse SNe (which will be addressed in \citealt{rosselli_forecast_2025}) and the color dependence of the \sn\ intrinsic scatter were not taken into account. The latter, in particular, was recently found to be the most important systematic in the measurement of the dark energy equation of state parameter from the latest analysis of the Dark Energy Survey (DES) \sn\ data \citep[hereafter V24]{vincenzi_dark_2024}. The measurement of $\fsig$ could be sensitive to correlations between color and velocity estimation or to possible non-Gaussianity introduced by the intrinsic scatter. To correct for potential biases that could arise from selection or color-dependent scattering while building the \sn\ Hubble diagram, most recent \sn\ cosmological analyses \citep{brout_pantheon_2022,des_collaboration_dark_2024} have relied on the BEAMS \citep{kunz_bayesian_2007} with Bias Corrections (BBC) framework \citep{kessler_correcting_2017}. In the BBC framework, large realistic simulations are used to compute a correction function of the biases of the Hubble diagram. This method has not yet been tested in the context of the $\fsig$ measurement.

In this paper, we present the first application of the BBC framework to simulations of the LSST 10 year survey
\sn\ sample in order to study the impact of intrinsic scatter systematics on the $\fsig$ constraint.
In Sect.~\ref{sec:sim} we describe our simulations. In Sect.~\ref{sec:method} we present the method used to infer PVs from Hubble diagram residuals and to constrain $\fsig$. Our results are presented and discussed in Sect.~\ref{sec:res}. Finally, we conclude in Sect.~\ref{sec:conc}.



\section{LSST Simulations}\label{sec:sim}
In the previous work of \citetalias{carreres_growth-rate_2023}, simulations were run using a dark matter halo catalog as \sn\ hosts and no direct correlations were introduced between \sns\ and their host galaxies. This could be problematic if such correlation introduces bias in the $\fsig$ analysis, for instance, \sn\ host galaxies are known to be more massive and hence could have a biased velocity distribution. On a smaller scale, more massive galaxies are also brighter, increasing the noise on \sn\ flux measurement and thus on the estimated distances, which could also bias the velocity distribution. In this work we aim to simulate the peculiar velocity field at the galaxy level along with \sns-host correlations.

\subsection{Peculiar velocity field and host galaxy simulation}
To produce the peculiar velocity field, the simulations of LSST \sns\ are generated on top of the Uchuu \texttt{UniverseMachine} simulated galaxy catalog \citep{ishiyama_uchuu_2021, aung_uchuu-universe_2023}. The Uchuu simulation consists of $12800^3$ dark matter particles of mass $m_p=3.27\times10^8 M_\odot$ in a $L_\mathrm{Uchuu} = 2 \ \mathrm{Gpc}~h^{-1}$ side length box. The simulation is initialized at a redshift of $z=127$ using the Planck15 results \citep{planck_collaboration_planck_2016} as the fiducial cosmology; the cosmological parameters are summarized in Table~\ref{tab:cosmo}. As described in \cite{aung_uchuu-universe_2023}, the Uchuu \texttt{UniverseMachine} galaxy catalog was constructed from the Uchuu $N$-body simulation with the \texttt{UniverseMachine} algorithm \citep{behroozi_universemachine_2019}, which aims to forward model the relationship between halos and galaxy properties in order to match observed data. We make use of the various galaxy properties present in this catalog, such as stellar mass and star-formation rate (SFR), to correlate host and \sn\ properties. The Uchuu \texttt{UniverseMachine} data are available as 50 snapshots at different redshifts ranging from $z=14$ to $z=0$. In this work we use the lowest redshift snapshot at $z=0$.

We subdivide the $(2 \ \mathrm{Gpc}~h^{-1})^3$ box into 8 non-overlapping sub-boxes of side length $L= 1 \ \mathrm{Gpc}~h^{-1}$ from which we produce 8 realizations of the 10 year LSST survey. We place the coordinate origin at the center of each box and convert distances to redshifts by numerically inverting the comoving distance-redshift relationship within the fiducial cosmology. The volume of each box corresponds to an effective redshift range of $z \in [0, 0.175]$. This redshift range is sufficient to study the LSST \sn\ PV survey, as the noise of PV measurements increase with redshift.

In order to simulate the noise in \sn\ flux measurements due to the luminosity of the host galaxies, the magnitude and Sérsic profile parameters of the galaxies are interpolated in the $(\log M_\mathrm{stellar} / M_\odot, \log \mathrm{SFR})$ parameter space from the galaxies produced for the \texttt{OpenUniverse} LSST-Roman simulations \citep{openuniverse_openuniverse2024_2025}. The Sérsic parameters are also used to generate each \sn\ position relative to its host galaxy with a probability that depends on its profile.

\begin{figure}
    \centering
    \includegraphics[width=\columnwidth]{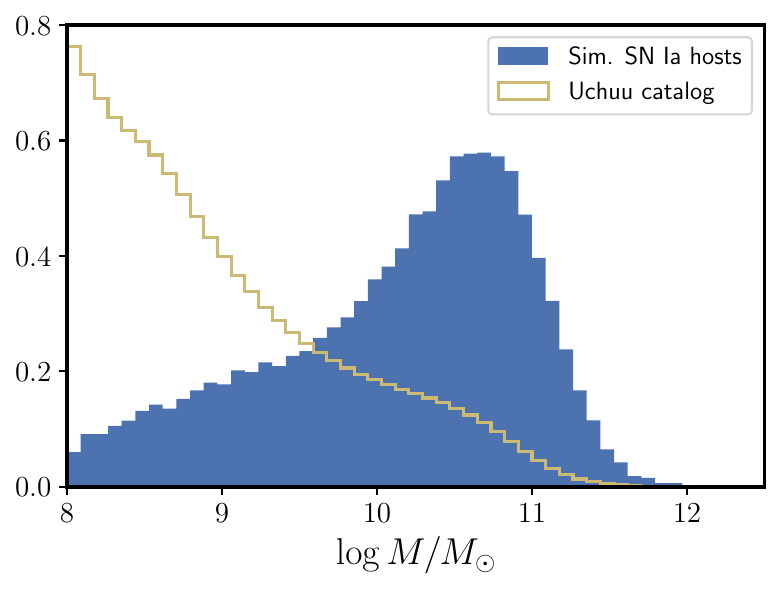}
    \caption{Distribution of host masses in the Uchuu \texttt{UniverseMachine} catalog (yellow) and of the \sn\ host masses from our LSST simulation (blue).}
    \label{fig:HostMass}
\end{figure}

As it was done in \cite{peterson_improving_2025}, the \sn\ hosts are randomly drawn according to the mass-dependent SNe~Ia rate described in Sect.~4 of \cite{wiseman_rates_2021}. In Fig.~\ref{fig:HostMass} we show the distribution of the masses of galaxies in the Uchuu \texttt{UniverseMachine} catalog compared to the distribution of host masses obtained in our simulation after applying the mass dependent random draw.

\begin{table}
    \caption{Fiducial cosmology parameters used in the simulation. These parameters are those from \cite{planck_collaboration_planck_2016}.}\label{tab:cosmo}   
    \centering
    \renewcommand{\arraystretch}{1.2}
    \begin{tabular}{cccccc}
    \hline\hline
    $h$ & $\Omega_m$ & $\Omega_b$ & $n_s$ & $\sigma_8$ & $\fsig$ \\[0.5ex]
    \hline
    0.6774 & 0.3089 & 0.0486 & 0.9667 & 0.8159 & 0.4253\\
    \hline
    \end{tabular}
\end{table}

\subsection{SN light-curve simulation and fitting}
\begin{figure}
    \centering
    \includegraphics[width=\columnwidth]{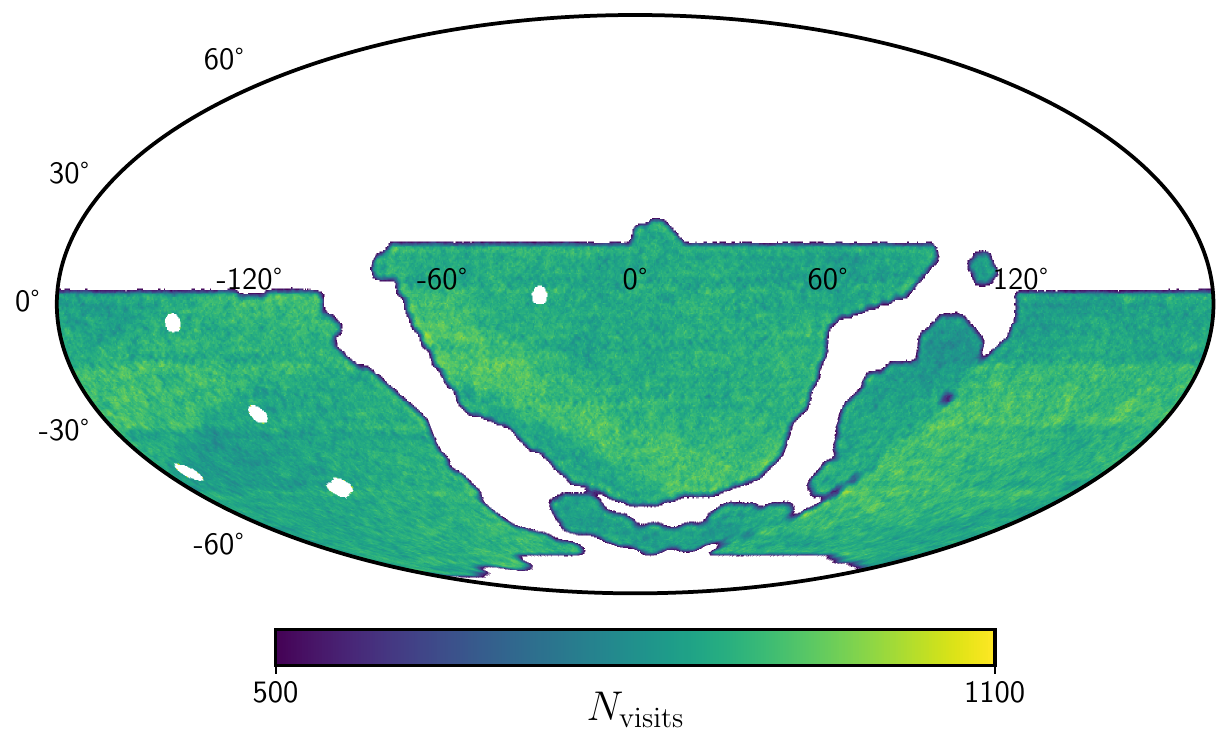}
    \caption{\texttt{HEALPix} pixels representation of the LSST survey WFD program. Pixels that contain less than 500 and more than 1100 visits are cut. \texttt{HEALPix} pixels are then sampled by \texttt{OpSimSummaryV2} to build the \texttt{SNANA} simulation input.}\label{fig:hp_view}
\end{figure}

\begin{figure*}
    \centering
    \includegraphics[width=\textwidth]{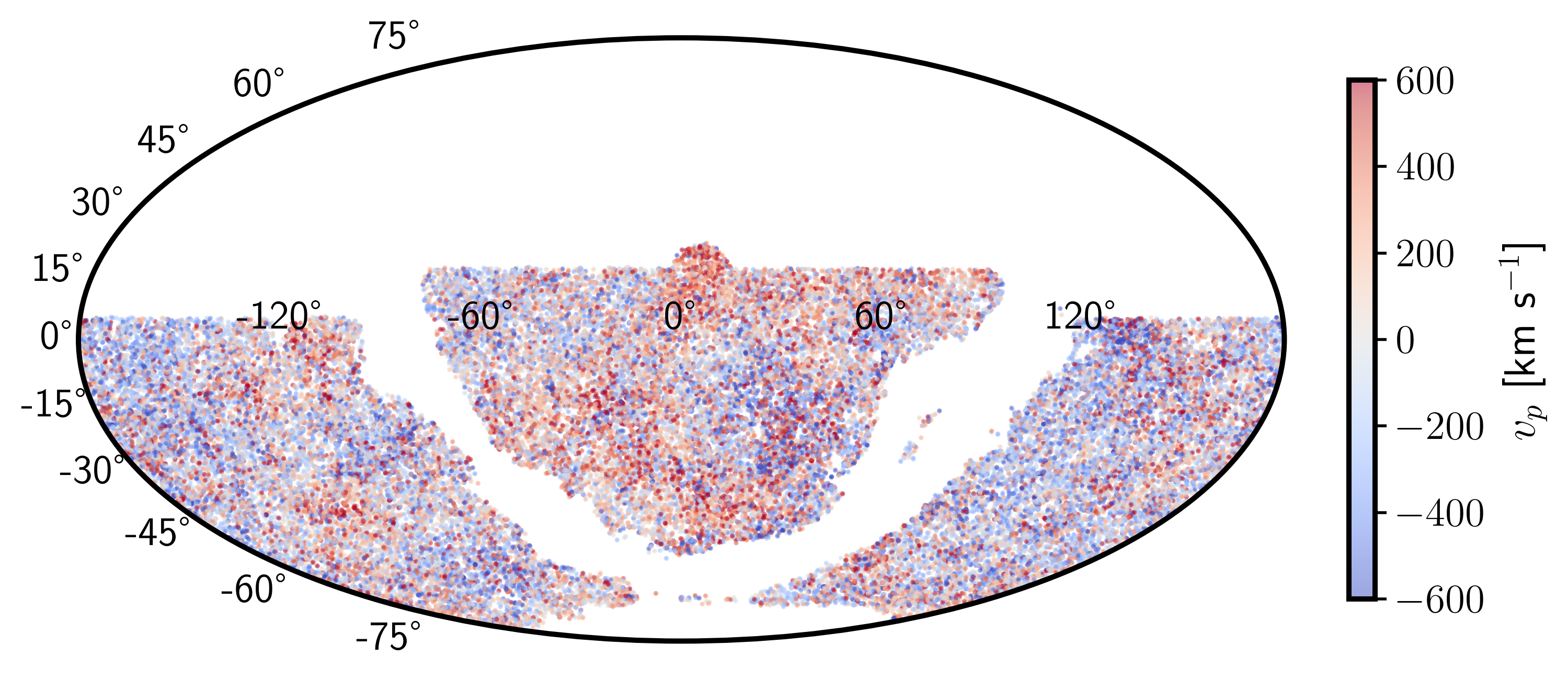}
    \caption{Angular distribution of the simulated \sn\ hosts. The colormap represents their peculiar velocities on the line of sight.}
    \label{fig:sn_loc}
\end{figure*}

The LSST survey will use the Simonyi Survey Telescope at Rubin Observatory that collects light with a 8.4~meter primary mirror. The Rubin Observatory LSST Camera has a $\sim$~9.4~deg$^2$ field-of-view that will allow the LSST survey to observe the entire Southern Hemisphere sky multiple times across 10 years.
We generate and analyze \sn\ data-like light-curves with the \texttt{SNANA} \git{RickKessler/SNANA}{} software library \citep{kessler_snana_2009} run through the \texttt{Pippin} \git{dessn/Pippin}{} pipeline \citep{hinton_pippin_2020}. We simulate \sn\ data equivalent to 10 years of the LSST survey.
To produce a realistic simulation of the LSST survey, \texttt{SNANA} requires two main inputs: a \texttt{SIMLIB} file that contains the observation characteristics and a \texttt{HOSTLIB} file that includes all the potential \sn\ host galaxies. These files are produced using the DESC software \texttt{OpSimSummaryV2} \git{LSSTDESC/OpSimSummaryV2}{}\citep{yoachim_lsstrubin_sim_2023}, that matches simulated observations made with the LSST software \texttt{OpSim} \git{lsst/rubin_sim}{}, with galaxy catalogs from Uchuu mocks. In this work we use the output \texttt{baseline\_v3.3\_10years} of \texttt{OpSim}. We focus on the Wide Fast Deep (WFD) program of the LSST survey that will observe in six filters over a large area of the southern sky. To build our simulation input files, \texttt{OpSimSummaryV2} represents the LSST WFD  survey as a collection of equal-area pixels computed using the \texttt{HEALPix}\footnote{\url{ http://healpix.sf.net}.} \citep{gorski_healpix_2005} scheme as implemented in the python library \texttt{healpy} \git{healpy/healpy}{} \citep{zonca_healpy_2019}. In this work, we use pixels with an area of $\sim0.05\text{ deg}^2$ and cut pixels with less than 500 observations over the 10 years of the survey. Then, \texttt{OpSimSummaryV2} associates to each of the pixels the LSST observations and Uchuu \texttt{UniverseMachine} host galaxies that lie within the radius of the Rubin Observatory field-of-view. These pixels are shown in Fig.~\ref{fig:hp_view}. The holes in the footprint correspond to the LSST Deep Drilling Field program that observes a small area with long exposures at a high cadence. These pixels were removed by imposing the cut that a pixel should be associated with less than 1100 observations. We note that we still have observations at these coordinates in our simulations, since observations are taken within the LSST field radius around each pixel. Finally, 50~000 pixels are randomly drawn and their associated observations and hosts are taken to build the \texttt{SNANA} simulation inputs. Fig.~\ref{fig:sn_loc} shows the angular distribution of the simulated \sns. 

To simulate a realistic number of \sns\, we use the \sn\ rate value from \cite{frohmaier_volumetric_2019}, which is $r_{v, \mathrm{F19}} = 2.27\times(1+z)^{1.7}\times10^{-5}$~\sn/Mpc/year and rescale it to our fiducial value of $h$ using
\begin{equation}
    r_{v,\mathrm{sim}} = r_{v, \mathrm{F19}}\times\left(\frac{0.7}{h_\mathrm{fid}}\right)^3.
\end{equation}

The simulated photometry is obtained by the integration of a spectral template of each \sn. In this work, we use the spectral template from the SALT3 model \citep{guy_salt2_2007, kenworthy_salt3_2021} as trained in \citetalias{vincenzi_dark_2024}. The SALT3 model gives a spectral template for \sns\ that depends on three parameters: the magnitude of the \sn\ $m_b$, the stretch $x_1$, and the color $c$.
The absolute magnitude of each \sn\ is defined by following the Tripp equation \citep{tripp_two-parameter_1998} as
\begin{equation}\label{eq:tripp}
    \boldsymbol{M_{b, \mathrm{sim}}^*} = M_{b, \mathrm{fid}} - \alpha_\mathrm{fid} \boldsymbol{x}_{\mathbf{1}, \mathrm{sim}} + \beta_\mathrm{fid} \boldsymbol{c}_\mathrm{sim} +  \boldsymbol{\Delta_M}(\boldsymbol{M_\mathrm{host}}, \gamma_\mathrm{fid}),
\end{equation}
where $\alpha_\mathrm{fid}$ is the stretch linear coefficient, $\beta_\mathrm{fid}$ is the color linear coefficient and $\boldsymbol{\Delta_M}$ is the mass-step function that we define below.
The observed magnitude is then computed as
\begin{equation}\label{eq:simmag}
    \boldsymbol{m_{b, \mathrm{sim}}}  = \boldsymbol{M_{b, \mathrm{sim}}^*}+ \boldsymbol{\mu}(\boldsymbol{z_\mathrm{cos}}) + 10\log_{10}\left(1+\boldsymbol{z_p}\right),
\end{equation}
where $\mu(\boldsymbol{z_\mathrm{cos}})$ is the distance moduli at the redshift of the \sn\ host and the last term is the relativistic beaming from peculiar velocity \citep{davis_effect_2011}. In Eqs.~\ref{eq:tripp} and~\ref{eq:simmag} we put in bold the quantities that differ for each \sn.

We generate simulations using four different SN Ia intrinsic scatter models. The first is the least realistic and is used for comparison: coherent (achromatic) Gaussian scatter with a standard deviation of $\sigma_M = 0.12$. We refer to this simulation as \simu{COH}. The next two are models that have been used historically in \sn\ analyses: the G10 model \citep{guy_supernova_2010} where 70\% of the luminosity scatter is due to achromatic variations and 30\% to chromatic ones, and the C11 model \citep{chotard_reddening_2011} where 25\% of the scattering is achromatic and 70\% is chromatic. These two models directly add a wavelength-dependent scatter to the spectral model provided by SALT3. We refer to these two simulations as \simu{G10} and \simu{C11}, respectively. Recent studies, however, have disfavoured these two models \citepalias{vincenzi_dark_2024}. A more realistic model is presented in \cite{brout_its_2021} (hereafter referred as BS21), which proposes that the scattering is caused by variation in dust extinction parameters of \sn\ host galaxies. We simulate the \citetalias{brout_its_2021} model using the dust parameters fitted in \cite{popovic_pantheon_2023} (hereafter referred as P23). We refer to this simulation as \simu{P23}.

In the simulations \simu{COH}, \simu{G10} and \simu{C11}, the values of $\alpha_\mathrm{fid}$ and $\beta_\mathrm{fid}$ are fixed to 0.15 and 3.1, respectively. The mass-step is defined by the function
\begin{equation}
    \Delta_M  =  
    \left\{\begin{matrix}
  - \gamma_\mathrm{fid}/2 \ \mathrm{if } \log M / M_\odot > 10\\
  + \gamma_\mathrm{fid}/2 \ \mathrm{if } \log M / M_\odot < 10
    \end{matrix}\right. ,
\end{equation}
with a value for $\gamma$ of 0.05~mag. The distribution of $x_1$ and $c$ are also chosen to be identical for these models. We use the parent populations $\mathcal{P}(x_1 | M_\mathrm{host})$ and $\mathcal{P}(c| M_\mathrm{host})$ derived in Table~4 and 8 of \cite{popovic_improved_2021} (hereafter referred as P21) which aim to model the correlations between \sn\ parameters and host masses.


In the \citetalias{brout_its_2021} model, the Tripp relation of Eq.~\ref{eq:tripp} is slightly modified such as
\begin{equation}
    \begin{split}
            \boldsymbol{M_{b,\mathrm{sim}}^*} =& M_{b, \mathrm{sim}} - \alpha_\mathrm{fid} \boldsymbol{x}_{\mathbf{1}, \mathrm{sim}} + \boldsymbol{\beta}_\mathrm{sim}\boldsymbol{c}_\mathrm{sim} \\
            &+  (\boldsymbol{R}_{\boldsymbol{V},\mathrm{sim}} + 1)\boldsymbol{E}_{\boldsymbol{\mathrm{dust}}, \mathrm{sim}},
    \end{split}
\end{equation}
where $\alpha_\mathrm{fid}$ is again fixed at a value of 0.15 and $\mathcal{P}(x_1 | M_\mathrm{host})$ is described by the same parent population as for the \simu{COH}, \simu{G10} and \simu{C11} simulations. The \citetalias{brout_its_2021} model assumes that the SN color can be divided into two components. The first corresponds to an intrinsic color variation, $\beta c$, where $\beta$ is a random variable following a Gaussian law $\mathcal{P}(\beta) \sim \mathcal{N}(\mu_\beta, \sigma_\beta)$ of parameters
\begin{equation}\label{eq:beta_p23}
    (\mu_\beta, \sigma_\beta) = (2.07, 0.22),
\end{equation}
and the intrinsic color parameter, $c$, follows a Gaussian random law $\mathcal{P}(c) \sim \mathcal{N}(\mu_c, \sigma_c)$ with parameters
\begin{equation}\label{eq:c_p23}
    (\mu_c, \sigma_c) = (-0.07, 0.05).
\end{equation}
The second component $(R_V + 1)E_\mathrm{dust}$ describes the effect of dust from the \sn\ host where the extinction ratio, $R_V$, is a Gaussian that depends on host mass $\mathcal{P}(R_V|M_\mathrm{host})\sim\mathcal{N}(\mu_{R_V}, \sigma_{R_V})$ with the parameters
\begin{equation}\label{eq:rvp23}
    (\mu_{R_V}, \sigma_{R_V}) = 
    \left\{\begin{matrix}
        (1.66, 0.95) & \text{ if } \log M/M_\odot > 10
        \\
        (3.25, 0.93)&\text{ else }
    \end{matrix}\right.,
\end{equation}
and the color excess due to dust, $E_\mathrm{dust}$, is described by an exponential distribution that depends on host masses $\mathcal{P}(E_\mathrm{dust}|M_\mathrm{host}, z)\sim\exp(\tau_{E_\mathrm{dust}})$ with parameters 
\begin{equation}\label{eq:taup23}
    \tau_{E_\mathrm{dust}} = 
    \left\{\begin{matrix}
        \left\{
            \begin{matrix}
                0.11 &\text{ if } \log M/M_\odot > 10 \\
                0.14 &\text{ else}
            \end{matrix}\right. &\ \text{ if }z<0.1\\[2ex]
        \left\{
            \begin{matrix}
                0.15 &\text{ if } \log M/M_\odot > 10 \\
                0.12 &\text{ else}
            \end{matrix}\right. &\ \text{ else }
    \end{matrix}\right..
\end{equation}
These values were derived using the method presented in \citetalias{popovic_pantheon_2023} and are presented in Table~3 of \citetalias{vincenzi_dark_2024}. We summarize the parameters for our four intrinsic scatter models in Table~\ref{tab:simpar}.

\begin{table*}
    \centering
    \caption{Summary of the \sn\ simulation parameters used for the \simu{COH}, \simu{G10}, \simu{C11} and \simu{P23} simulations.\label{tab:simpar}}
    \renewcommand{\arraystretch}{1.5}
    \begin{tabularx}{\textwidth}{c@{\extracolsep{\fill}}||ccccccc}
        \textbf{Simulations}~ & $\alpha$ & $\beta$ & $\gamma$ & $\mathcal{P}(x_1 | M_\mathrm{host})$ & $\mathcal{P}(c | M_\mathrm{host})$ & $\mathcal{P}(R_V|M_\mathrm{host})$ & $\mathcal{P}(E_\mathrm{dust}|M_\mathrm{host}, z)$\\[1ex]
        \hline\hline
        \simu{COH}, \simu{G10} & 0.15  & 3.1  & 0.05   & Table~4 of \citetalias{popovic_improved_2021} &Table~8 of \citetalias{popovic_improved_2021} & - & - \\
        \simu{C11} & 0.15  & 3.8 & 0.05 & Table~4 of \citetalias{popovic_improved_2021} & Table~8 of \citetalias{popovic_improved_2021} & - & -  \\
        \simu{P23}                         & 0.15  & $\mathcal{N}(\mu_\beta,\sigma_\beta)$ & - & Table~4 of \citetalias{popovic_improved_2021} & $\mathcal{N}(\mu_c,\sigma_c)$ & Eq.~\ref{eq:rvp23} & Eq.~\ref{eq:taup23}\\[1ex]
        \hline
    \end{tabularx}    
    
\end{table*}

We include in our simulation the reddening effect of the Milky Way dust by using the extinction model of \cite{fitzpatrick_correcting_1999} and the Milky Way reddening map from \cite{schlafly_measuring_2011}.

During the simulation process, we apply the magnitude-dependent detection efficiency of the LSST survey as computed in \cite{sanchez_snia_2022} to each observation. A \sn\ is considered detected if we have at least one observation that passes this detection efficiency.
To discard poor-quality \sn\ and to further limit the number of simulated \sns\ stored, we apply different quality cuts at the end of the simulation stage. We require that the light-curve has at least five observations with a signal-to-noise ratio (SNR) above five and that at least two of them are in different bands amongs the $griz$ bands. In addition, we require at least one observation in the time range of $-20$ to $-5$ days before the \sn\ peak brightness time and one in the time range of 5 days to 40 days after this peak brightness time. To mitigate the effect of dust reddening, we cut all \sns\ that are observed through the plane of the Milky Way by removing events with a Milky Way dust extinction $E(B-V)$ above 0.25~mag.

After the simulation, the light-curves are fitted with the same SALT3 model that was used to simulate them. The fitted parameters are the time of peak $t_0$, the stretch $x_1$, the color $c$ and the normalization factor $x_0$ that is related to magnitude as
\begin{equation}
    m_B = -2.5\log x_0 + C,
\end{equation}
where $C$ is a constant offset.
We also obtain the covariance matrix of the fitted parameters for each \sn
\begin{equation}
    \text{C}_\mathrm{SALT} = 
    \begin{pmatrix}
       \sigma_{m_b}^2 & \text{Cov}[m_b, x_1]  & \text{Cov}[m_b, c] \\
       \text{Cov}[m_b, x_1] & \sigma_{x_1}^2   & \text{Cov}[x_1, c] \\
        \text{Cov}[m_b, c] & \text{Cov}[x_1, c]  & \sigma_c^2
    \end{pmatrix}.
\end{equation}

\section{Methods}\label{sec:method}
We present here the two approaches used to build the Hubble diagram which we compare in the context of the $\fsig$ constraint. The simple framework refers to the approach that was used in \citetalias{carreres_growth-rate_2023}. The BBC framework refers to the framework originally presented in \cite{kessler_correcting_2017} and improved in \citetalias{popovic_improved_2021}.

\subsection{Simple framework}
In our simple framework, the Hubble residuals are written
\begin{equation}
     \boldsymbol{\Delta\mu} = \boldsymbol{\mu_\mathrm{obs}} - \mu_\mathrm{cos}(\boldsymbol{z_\mathrm{obs}}),
\end{equation}
where $\boldsymbol{\mu_\mathrm{obs}}$ is computed from SALT fitted parameters as
\begin{equation}
    \boldsymbol{\mu_\mathrm{obs}} = \boldsymbol{m_b} - \left(M_0 - \alpha \boldsymbol{x_1} + \beta \boldsymbol{c}  + \boldsymbol{\Delta_M}\left(\mathbf{M_\mathrm{host}},\gamma\right)\right),
\end{equation}
with $\alpha$, $\beta$, $M_0$ and $\gamma$ the free nuisance parameters that need to be fitted. 
The cosmological distance modulus $\mu_\mathrm{cos}(\boldsymbol{z_\mathrm{obs}})$ depends on cosmological parameters that are usually fitted in the cosmological analysis. However, at low-$z$ the Hubble diagram does not strongly depend on the cosmological parameters and in this framework we instead obtain these parameters using the fiducial cosmology of the simulation to compute the distance modulus as
\begin{equation}
    \mu_\mathrm{cos}(z) = 5\log\left((1 + z) r(z)\right) + 25,
\end{equation}
where the comoving distance $r(z)$ in Mpc is computed as
\begin{equation}
    r(z) = \frac{c}{H_0}\int_0^z \frac{1}{\sqrt{\Omega_m(1+z)^3+ 1 - \Omega_m}},
\end{equation}
with $H_0$ and $\Omega_m$ respectively the Hubble constant and the matter energy density at $z=0$.
The covariance matrix of $\mathbf{\Delta\mu}$ is given by
\begin{equation}
    C^{\mu\mu} = \text{diag}\left[\boldsymbol{\sigma_\mathrm{SALT}}^2\right]+\sigma_\mathrm{int}^2 \mathbf{I},
\end{equation}
where the SALT error term is computed for each \sn\ as
\begin{equation}
    \sigma_\mathrm{SALT}^2 = A \text{C}_\mathrm{SALT} A^T,
\end{equation}
with $A= (1, \alpha,  -\beta)$, and $\sigma_\mathrm{int}$ is an additional nuisance parameter that corresponds to an achromatic scatter which must be fitted.

\subsection{The BBC framework}\label{sec:method:bbc}
In the BBC framework, large survey-specific simulations are used to correct the estimated distance modulus as a function of \sn\ parameters. 
Since the bias correction simulation is computationally intensive, as it is roughly 40 times the full LSST simulation size, we produce only one for each intrinsic scatter model to be used for all of our eight mocks. In order to not interfere with the PV statistic when applying the bias correction, we set the PVs of the bias correction simulation to zero. The population parameters for $x_1$ and $c$ used to produce the bias correction simulations are identical to the ones used for our baseline simulations.

For the \simu{COH}, \simu{G10} and \simu{C11} simulations we follow the approach described in \citetalias{popovic_improved_2021} as BBC-7D.  In this method, the parameters of the \sns\ from the large simulation are distributed over a 7D grid $\left\{z_\mathrm{obs}, x_1, c, \alpha, \beta, \theta, M_\mathrm{host}\right\}$. Thus, the bias correction simulations are generated over a $2\times 2$ grid of the parameters $\alpha$ and $\beta$.
The extra parameter $\theta$ is a magnitude shift of $\pm 0.06$~mag introduced in the bias correction simulation as prescribed in \citetalias{popovic_improved_2021}. It helps to take into account bias in the recovered $\gamma$ value emerging from correlations between host masses and the SALT stretch and color parameters \citep{smith_first_2020}. Then, bias corrections are computed by averaging in each cell the difference between the estimated SALT parameters and their simulated value,
\begin{equation}
    \delta_{p, \mathrm{cell}} = \langle p - p_\mathrm{sim} \rangle_\mathrm{cell}.
\end{equation}
The bias correction function $\delta_p(z_\mathrm{obs}, x_1, c, \alpha, \beta, \theta, M_\mathrm{host})$ is then obtained by a linear interpolation over the cells.
Thus, the corrected \sn\ distance moduli are given by
\begin{equation}
    \begin{split}
            \boldsymbol{\mu_{\mathrm{obs}, \mathrm{BBC}}} =& (\boldsymbol{m_b} - \delta_{m_b}) \\
            &-\left(M_{0,z_i} - \alpha (\boldsymbol{x_1} - \delta_{x_1}) + \beta(\boldsymbol{c} - \delta_{c})\right. \\ &+\left.\boldsymbol{\Delta_M}\left(\mathbf{M_\mathrm{host}},\gamma\right) \right),
    \end{split}
\end{equation}
where the $M_{0,z_i}$ are offsets fitted in each redshift bin. 

For the \simu{P23} simulation we use the approach called BBC-BS20 in \citetalias{popovic_improved_2021} where the parameters of the grid are reduced to $\left\{z_\mathrm{obs}, x_1, c, M_\mathrm{host}\right\}$, since $\beta$ has a different value for each SNe and the mass-step is a consequence of the effect of dust on color. The bias correction is directly computed on the difference between the fiducial distance modulus from simulated cosmology and the recovered one
\begin{equation}\delta\mu_{\mathrm{bias}, \mathrm{cell}} = \langle \mu_\mathrm{obs} - \mu_\mathrm{fid}\rangle_\mathrm{cell}.
\end{equation}
Similarly as in BBC-7D, the bias correction function $\delta\mu_\mathrm{bias}(z, x_1, c, M_\mathrm{host})$ is linearly interpolated over the cells and the distance modulus is given by
\begin{equation}
    \boldsymbol{\mu_{\mathrm{obs}, \mathrm{BBC}}} = \boldsymbol{m_b} - \left(M_{0,z_i} - \alpha \boldsymbol{x_1} + \beta \boldsymbol{c}\right) - \boldsymbol{\delta\mu_\mathrm{bias}}.
\end{equation}
The corresponding errors are computed as
\begin{equation}
    \boldsymbol{\sigma_{\mu,\mathrm{BBC}}}^2 = \boldsymbol{\sigma_\mathrm{SALT}}^2 + \sigma_\mathrm{int}^2.
\end{equation}
Using the bias correction functions, the nuisance parameters $\alpha$, $\beta$, $\gamma$ and $M_{0, z_i}$ along with the intrinsic scatter $\sigma_\mathrm{int}$ are determined by the fitting procedure described in \cite{marriner_more_2011}, \cite{kessler_correcting_2017} and \cite{kessler_binning_2023}. In this procedure, the residuals to the fiducial cosmology of the bias correction simulations,
\begin{equation}
    \textbf{HR}=  \boldsymbol{\mu_{\mathrm{obs}, \mathrm{BBC}}} - \boldsymbol{\mu}_\mathrm{fid}(\boldsymbol{z_\mathrm{obs}}),
\end{equation}
are minimized with respect to the nuisance parameters using $\chi^2$ minimization method.

The Hubble diagram residuals are then obtained as
\begin{equation}    \boldsymbol{\Delta\mu_\mathrm{BBC}} = \boldsymbol{\mu_{\mathrm{obs}, \mathrm{BBC}}} - \left[\boldsymbol{\mu_\mathrm{cos}} (\boldsymbol{z_\mathrm{obs}})+ M_0\right],
\end{equation}
where $M_0$ is a global offset of the Hubble diagram. We note that in this work we used the same cosmology for our data simulations and bias correction simulations. Thus the expected value of $M_0$ is $M_0=0$.
The associated covariance matrix is 
\begin{equation}
    C^{\mu\mu} = \text{diag}\left[\boldsymbol{\sigma_{\mu, \mathrm{BBC}}}^2\right],
\end{equation}
where $\sigma_{\Delta\mu, \mathrm{BBC}}$ is computed for each \sn\ as indicated in Eq.~3 of \cite{kessler_correcting_2017}.

In this work we also test the inclusion of the systematic covariance matrix due to the parametrization of the \citetalias{brout_its_2021} intrinsic scatter model. To compute this matrix in the BBC framework we generated bias correction simulations to apply to our \simu{P23} data simulation using four variations of our nominal \citetalias{brout_its_2021} model parameters. As in \citetalias{vincenzi_dark_2024} we use the parameter values originally found in \citetalias{brout_its_2021} and the three random realizations of the parameters drawn from the posterior distributions that were computed in \citetalias{popovic_pantheon_2023}. The covariance matrix is computed following the method presented in \cite{conley_supernova_2010} as 

\begin{equation}
    \text{C}_{ij}^{\mu\mu, \mathrm{int. scat.}} = \Delta\mu_i^\mathrm{BS21}\Delta\mu_j^\mathrm{BS21} + \frac{1}{3}\sum_{p=1}^3\Delta\mu_i^{\mathrm{P23},p}\Delta\mu_j^{\mathrm{P23},p},
\end{equation}
where the $\Delta\mu^V_i$ are the differences between distance moduli obtained using the nominal bias correction simulations and the ones obtained using the variation $V$ of the \citetalias{brout_its_2021} model parameters. The total covariance is then given by
\begin{equation}
    C^{\mu\mu} =  \text{diag}\left[\boldsymbol{\sigma_{\mu, \mathrm{BBC}}}^2\right] + \text{C}_{ij}^{\mu\mu, \mathrm{int. scat.}}.
\end{equation}

\subsection{Sample cuts}\label{sec:method:samplecuts}
Before the Hubble diagram construction we apply quality cuts to our sample of \sns\ in order to remove outliers. We impose cuts on SALT parameters; the stretch parameter must be $|x_1|<3$, the color has to be $|c|<0.3$. The error on the time of peak luminosity $t_0$ must be $\sigma_{t_0}<2$ and the one on the stretch parameter must be $\sigma_{x_1} < 1$. We also discard potential bad light-curve fits by requiring that the fit probability is $\mathcal{P}_\mathrm{SALTfit} > 0.001$\footnote{The probability of the fit is defined as $\mathcal{P}_\mathrm{SALTfit}=\int_{\chi^2}^\infty \mathcal{P}_{\chi^2}(x;k)dx$ where $\mathcal{P}_{\chi^2}(x;k)$ is the chi-square distribution function with $k$ degrees of freedom.}. During the BBC fit procedure, we iteratively reject outlier events by removing those that have a $\chi^2_\mathrm{HD} > 16$\footnote{$\chi_\mathrm{HD}^2 = \left(\mu_\textrm{obs} - \mu(z_\mathrm{obs})\right)^2/\sigma_\mu^2$}, equivalent to a $4\sigma$ clipping.

To include a \sn\ in the fit of $\fsig$ we need to be able to estimate its host redshift. Hence, after the BBC Hubble diagram fit we require that the \sn\ has been matched to a host using a cut on the dimensionless distance $d_\mathrm{DLR} < 4$ where $d_\mathrm{DLR}$ is defined as the ratio between the angular separation of the \sn\ to its host and the radius of the host in the \sn\ direction \citep{gupta_host_2016,qu_dark_2024}. Additionally, in this paper we focus on the $\fsig$ fit in the redshift range $z\in [0.02, 0.1]$ (a more extensive forecast for different redshift ranges is provided in \citealt{rosselli_forecast_2025}) and therefore cut the \sns\ which fall outside this range. The lower redshift cut is implemented in order to avoid bias on the velocity estimator, as shown in Appendix~A.2 of \citetalias{carreres_growth-rate_2023}.
Averaged over our eight mocks, the final sample size is $\sim6600$ \sns\ and is summarized for each intrinsic scatter model simulation in Table~\ref{tab:meanN}.

\begin{table}[]
    \caption{Mean number of \sns\ after cuts over our 8 mocks for each of our intrinsic scatter model simulations in the redshift range $0.02<z<0.1$.}\label{tab:meanN}
    \renewcommand{\arraystretch}{1.5}
    \begin{tabular}{c|cccc}
         & \simu{COH} & \simu{G10} & \simu{C11} & \simu{P23} \\
         \hline 
       $\langle N_\mathrm{SN}\rangle$  & 6670 & 6658 & 6534 & 6917\\[1ex]
       \hline
    \end{tabular}
    \label{tab:my_label}
\end{table}

\subsection{Constraint on $\fsig$ using the maximum likelihood method}
We fit for $\fsig$ using the maximum likelihood method previously used in \citetalias{carreres_growth-rate_2023} and implemented in the Python library \texttt{flip} \git{corentinravoux/flip}{} \citep{ravoux_generalized_2025}.

\subsubsection{Velocity estimates and covariance matrix}
Line-of-sight velocities can be estimated from the Hubble diagram residuals using the transformation
\begin{equation}
    \boldsymbol{v} = J(\boldsymbol{z_\mathrm{obs}})\boldsymbol{\Delta\mu},
\end{equation}
where $J(\boldsymbol{z_\mathrm{obs}})$ is a diagonal matrix that depends on redshift:
\begin{equation}
    J(\boldsymbol{z_\mathrm{obs}}) = \mathrm{diag}\left[-\frac{c\ln10}{5}\left(\frac{(1+\boldsymbol{z_\mathrm{obs}})c}{H(\boldsymbol{z_\mathrm{obs}})r(\boldsymbol{z_\mathrm{obs}})} - 1\right)^{-1}\right],
\end{equation}
where $H(z_\mathrm{obs})$ is the Hubble parameter evaluated at the observed redshift.

The statistic of the line-of-sight velocities is computed using the velocity divergence $\theta$ linked to the velocity field by
\begin{equation}
    \nabla.\mathbf{v}(\mathbf{r}, a)= -aH(a)f(a)\theta(\mathbf{r}, a),
\end{equation}
where $a$ is the scale factor $a=1/(1+z)$.
In Fourier space this relation can be written as
\begin{equation}
    \mathbf{v}(\mathbf{k}, a) = -i aH(a)f(a)\frac{\mathbf{\hat{k}}}{k}\theta(\mathbf{k}, a).
\end{equation}
In the following we assume $a=1$ since we are working on the Uchuu \texttt{UniverseMachine} snapshot at $z=0$.

The covariance of line-of-sight velocities is given by
\begin{align}
        C^{vv}_{ij}&=\langle(\mathbf{v}_i.\hat{\mathbf{r}_i})((\mathbf{v}_j.\hat{\mathbf{r}_j})\rangle \\
         &=(aHf)^2 \iint \frac{d^3 \mathbf{k}_i}{(2\pi)^3}\frac{d^3 \mathbf{k}_j}{(2\pi)^3}\frac{\mu_i\mu_j}{k_ik_j}\langle\theta(k_i)\theta^*(k_j)\rangle,
\end{align}
where $\mu = \mathbf{\hat{k}}.\mathbf{\hat{r}}$ with $\hat{r}$ the normalized position vector.
The power spectrum of $\theta$ is defined such as $\langle\theta(k_i)\theta^*(k_j)\rangle=(2\pi)^3\delta_D(k_i-k_j)P_{\theta\theta}(k)$.
From the power spectrum, the velocity covariance matrix is computed as
\begin{equation}\label{eq:cvv}
    \text{C}_{ij}^{vv} = \frac{H_0^2}{2\pi^2}\frac{(\fsig)^2}{(\fsig)^2_{\rm fid}} \int_{k_\mathrm{min}}^{k_\mathrm{max}} f_{\rm fid}^2P_{\theta\theta}(k)W_{ij}(k; \mathbf{r}_i, \mathbf{r}_j) {\rm d}k.
\end{equation}
where $W_{ij}(k; \mathbf{r}_i, \mathbf{r}_j)$ is the window function such that
\begin{equation}
    \begin{split}
     W_{ij}(k; \mathbf{r}_i, \mathbf{r}_j) &= 
            \frac{1}{3} \left[ j_0\left(k r_{ij}\right) 
                             - 2j_2\left(k r_{ij}\right) 
                         \right] \cos (\alpha_{ij}) \\
            & \ \ + \frac{1}{r_{ij}^2} j_2\left(k r_{ij}\right)r_i r_j \sin^2(\alpha_{ij}),
    \end{split}
\end{equation}
with $r_{ij} = \left|\mathbf{r}_j - \mathbf{r}_j\right|$ the distance between two \sn\ hosts and $\alpha_{ij}$ the angle between the two lines of sight such that $\cos(\alpha_{ij}) = \mathbf{r}_i.\mathbf{r}_j / r_i r_j$. The integration lower bound $k_\mathrm{min}$ is set to correspond to the smallest modes of the $N$-body simulation, $k_\mathrm{min} = 2\pi / L_\mathrm{Uchuu} \sim 3\times10^{-3} \ \text{Mpc}^{-1}~h$. We take $k_\mathrm{max} = 0.2 \ \text{Mpc}^{-1}~h$ as the upper boundary since using a higher value of $k_\mathrm{max}$ does not significantly change the value of the covariance.

The linear power spectrum $P_{\theta\theta, \mathrm{lin}}(k)$ is computed using the \texttt{CLASS} library \git{lesgourg/class_public}{} \citep{lesgourgues_cosmic_2011}. We take into account smaller scale non-linearities using the formula derived in \cite{bel_accurate_2019},
\begin{equation}
    P_{\theta\theta}(k) =  P_{\theta\theta, \mathrm{lin}}(k) \text{exp}\left[-k\left(a_1+a_2 k+ a_3 k^2\right)\right],
\end{equation}
with the coefficient values
    \begin{equation}
        \begin{matrix}a_1 &=& -0.817 + 3.198 \sigma_8 \\
               a_2 &=& 0.877 - 4.191\sigma_8 \\
               a_3 &=& -1.199 + 4.629\sigma_8\end{matrix}.
    \end{equation}
These coefficients were fitted on a $N$-body simulation and the accuracy of the resulting power-spectrum is shown to be $>95\%$ for our range of $k$ in \cite{bel_accurate_2019}.

\subsubsection{The redshift space damping function and calibration of the $\sigma_u$ parameter}\label{sec:method:pwsigu}
Since positions of galaxies are evaluated using the observed redshift $z_\mathrm{obs}$, their distances are affected by velocities through redshift space distortions \citep{kaiser_clustering_1987}. To take into account this effect on the velocity covariance we need to introduce a damping function $D_u$. An empirical expression for $D_u$ has been proposed in \cite{koda_are_2014} and is expressed as
\begin{equation}
    D_u(k) = \frac{\sin\left(k\sigma_u\right)}{k\sigma_u},
\end{equation}
where $\sigma_u$ is a nuisance parameter that needs to be determined.
With the damping function the velocity covariance of Eq.~\ref{eq:cvv} is rewritten as
\begin{equation}\label{eq:cvvDu}
    \text{C}_{ij}^{vv} = \frac{H_0^2}{2\pi^2}\frac{(\fsig)^2}{(\fsig)^2_{\rm fid}} \int_{k_\mathrm{min}}^{k_\mathrm{max}} f_{\rm fid}^2P_{\theta\theta}(k)D_u^2(k) W_{ij}(k; \mathbf{r}_i, \mathbf{r}_j) {\rm d}k.
\end{equation}

\begin{figure}
    \centering
    \includegraphics[width=\linewidth]{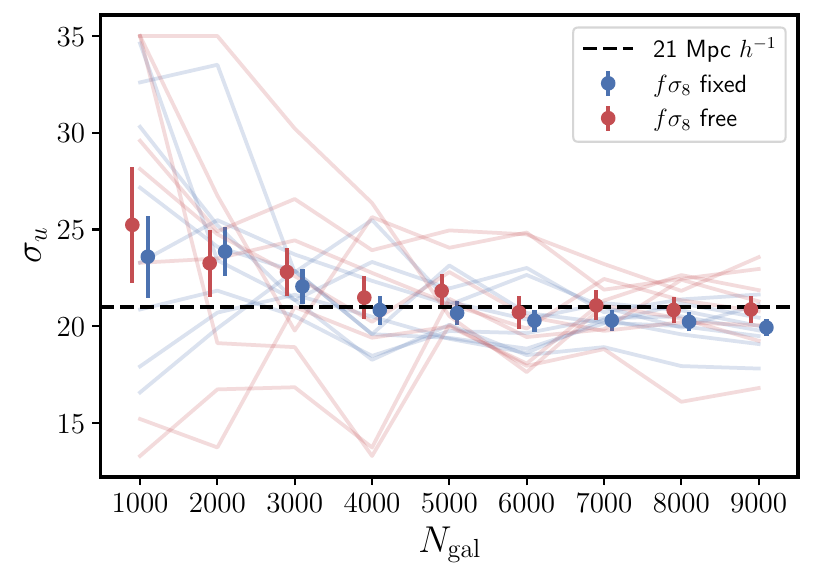}
    \caption{Results of the fit for $\sigma_u$ on true velocities from Uchuu simulated galaxies. The red points represent the result in the case where $\fsig$ is fixed to the fiducial value of 1. The blue points represent the case where $\fsig$ is fitted along with $\sigma_u$.}
    \label{fig:su_res}
\end{figure}

The covariance of the velocity is a computationally intensive task and fitting for the nuisance parameter of the redshift space distortion damping function, $\sigma_u$, requires us to either recompute the covariance on each iteration of the likelihood or to pre-compute the covariances for some given values of $\sigma_u$ and interpolate them. In the case of LSST, the number of \sns\ used to fit for $\fsig$ for the redshift range considered here is of the order of $\sim O(6600)$. Therefore, we choose to fix $\sigma_u$ to a pre-calibrated value. In order to determine the value of $\sigma_u$ we fit on the true peculiar velocities directly from the Uchuu catalog. We randomly draw 9000 galaxies in each of our eight mocks and then fit for $\sigma_u$ using two different variations of the fit. In the first we allow $\fsig$ to vary. In the other we fix $\fsig$ value to the fiducial one. The results are presented in Fig.~\ref{fig:su_res}.
We can see in Fig.~\ref{fig:su_res} that in the case where $\fsig$ is fitted along the value of $\sigma_u$ seems to converge towards $\sim 21$~Mpc~$h^{-1}$. In the case where $\fsig$ is fixed we can see that after converging around a value of $\sim 21$~Mpc~$h^{-1}$ the $\sigma_u$ parameter tends to slightly decrease. A possible interpretation is that with the increase of density, we start to probe more non-linear scales that impact the value of the damping factor. In this work we decide to fix $\sigma_u$ to the value of $\sim 21$~Mpc~$h^{-1}$. This is comparable to the values ranging from 19~Mpc~$h^{-1}$ to 23~Mpc~$h^{-1}$ found in \cite{lai_using_2023} but is slightly higher than the value of $\sim15$~Mpc~$h^{-1}$ found in \cite{koda_are_2014} and \citetalias{carreres_growth-rate_2023}. We further discuss this systematic in Sect.~\ref{sec:res:sigu}.

\subsubsection{The $\fsig$ likelihood}
The growth-rate likelihood is given by
\begin{equation}\label{eq:lik}
    \begin{split}
    \mathcal{L}(\Theta, \Theta_\mathrm{HD}) &= \left(2\pi\right)^{-\frac{n}{2}} \left|\text{C}(\Theta, \Theta_\mathrm{HD}) \right|^{-\frac{1}{2}} \\
    &\times \exp\left[-\frac{1}{2}\boldsymbol{v}^T(\Theta_\mathrm{HD})\text{C}^{-1}(\Theta, \Theta_\mathrm{HD})\boldsymbol{v}(\Theta_\mathrm{HD})\right],
    \end{split}
\end{equation}
where $\Theta = \left\{\fsig, \sigma_v, \sigma_u\right\}$ are the parameters linked to the PV field, $\Theta_\mathrm{HD} = \left\{\alpha, \beta, M_0, \sigma_\mathrm{int} \right\}$ are the parameters linked to the \sn\ standardization and $\text{C}(\Theta, \Theta_\mathrm{HD}) $ is the covariance matrix given by
\begin{equation}
   \text{C}(\Theta, \Theta_\mathrm{HD})  = \text{C}^{vv}(\fsig, \sigma_u) +  C^{{vv},\text{obs}}(\Theta_\mathrm{HD})  + \sigma_v^2 \mathbf{I},
\end{equation}
where the covariance of observations is given by
\begin{equation}
    C^{{vv},\text{obs}}(\Theta_\mathrm{HD}) = J \text{C}^{\mu\mu}(\Theta_\mathrm{HD}) J^T.
\end{equation}

In some cases two \sns\ are assigned to the same host. In this case the covariance matrix of velocities has two identical rows and is therefore not invertible. This issue is handled by a weighted-average of the velocities estimated with the \sn\ in the same galaxy $g$. We define the weights matrix $W$ with elements given by
\begin{equation}
    w_{ij} = 
        \left\{\begin{matrix}
            \frac{\sigma_{\hat{v}_j}^{-2}}{\sum_{p\in g_i}\sigma_{\hat{v}_{p}}^{-2}}& \text{ if } \text{SN}_j\in g_i
     \\
        0& \text{else}
                \end{matrix}\right.,
\end{equation}
The velocity data vector then becomes
\begin{equation}
    \mathbf{v} \rightarrow W\mathbf{v},
\end{equation}
and the associated covariance of observations is given by the transformation
\begin{equation}
    \text{C}^{{vv}, obs} \rightarrow W\text{C}^{{vv}, obs}W^T.
\end{equation}

We note that in the BBC framework all the $\Theta_\mathrm{HD}$ parameters are fixed prior to the $\fsig$ fit during the construction of the Hubble diagram. In the case of the simple framework, the $\Theta_\mathrm{HD}$ parameters are fit jointly with $\Theta$.
We perform the fit by minimizing the negative logarithm of the likelihood of Eq.~\ref{eq:lik} using the \texttt{Minuit} algorithm \citep{james_minuit_1975} as implemented in the Python library \texttt{iminuit} \citep{dembinski_scikit-hepiminuit_2024}.

\section{Results}\label{sec:res}
In this section, we present results obtained after minimizing the negative logarithm of the likelihood from Eq.~\ref{eq:lik}. We focus on estimated PVs and the recovered value of $\fsig$. Results obtained for the parameters of the Hubble diagram are presented in Appendix~\ref{app:hdpar}. 

\subsection{Estimation of PVs for the four scatter models}
\begin{figure*}
    \centering
    \includegraphics[width=0.45\textwidth]{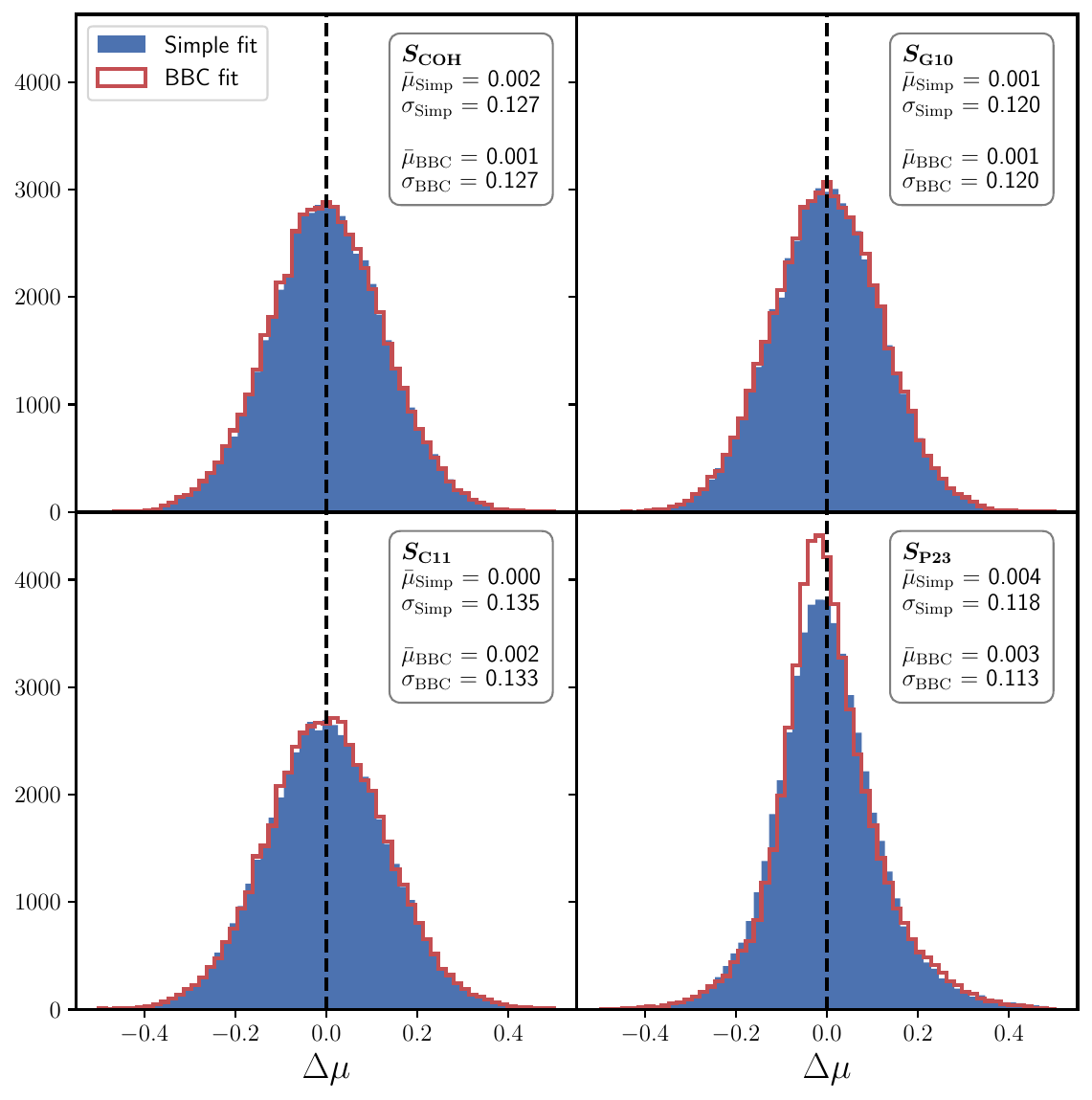} 
        \includegraphics[width=0.45\textwidth]{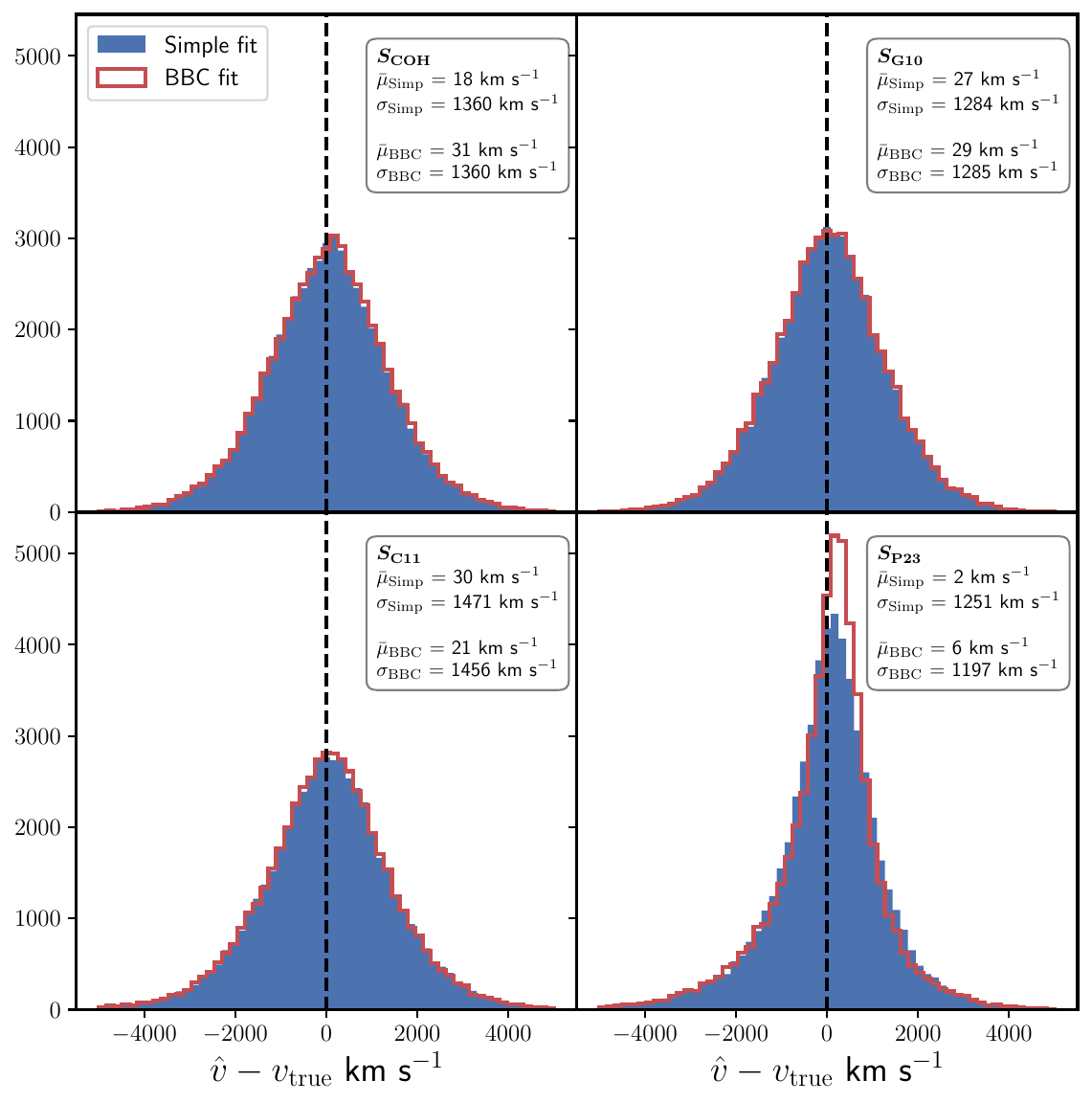} 
    \caption{Histograms of the Hubble diagram residuals (left) and velocity bias residuals (right). Results for the simple fit are in blue and results for the BBC fit are in red. We added in legend the standard deviation of the samples.}
    \label{fig:residuals}
\end{figure*}
In the maximum likelihood method, $\fsig$ is fitted as the scale of the covariance matrix of the PV distribution, thus it is important to examine this distribution prior to fitting. We represent the distributions of the Hubble diagram residuals and the difference between estimated velocities and the true ones for our four simulations stacked over the eight mocks in Fig.~\ref{fig:residuals}. The Hubble diagram scatter has a similar amplitude for the four models, ranging between $\sigma\sim0.11$~mag for \simu{P23} to $\sigma\sim 0.14$~mag for \simu{C11}. The two fitting methods yield the same scatter for the less chromatic models of \simu{COH} and \simu{G10}. For the chromatic models of \simu{C11} and \simu{P23}, the BBC fit 
yields a slightly but non-significant better scatter than the simple fit. These values of the scatter of the Hubble residuals result in estimated PV scatters in a range between $\sim1200$~km~s$^{-1}$ and $\sim1500$~km~s$^{-1}$.
We note that for \simu{P23}, the distribution of the Hubble diagram residuals deviates from the Gaussian one and seems positively skewed.


\begin{figure*}
    \centering
    \includegraphics[width=0.45\textwidth]{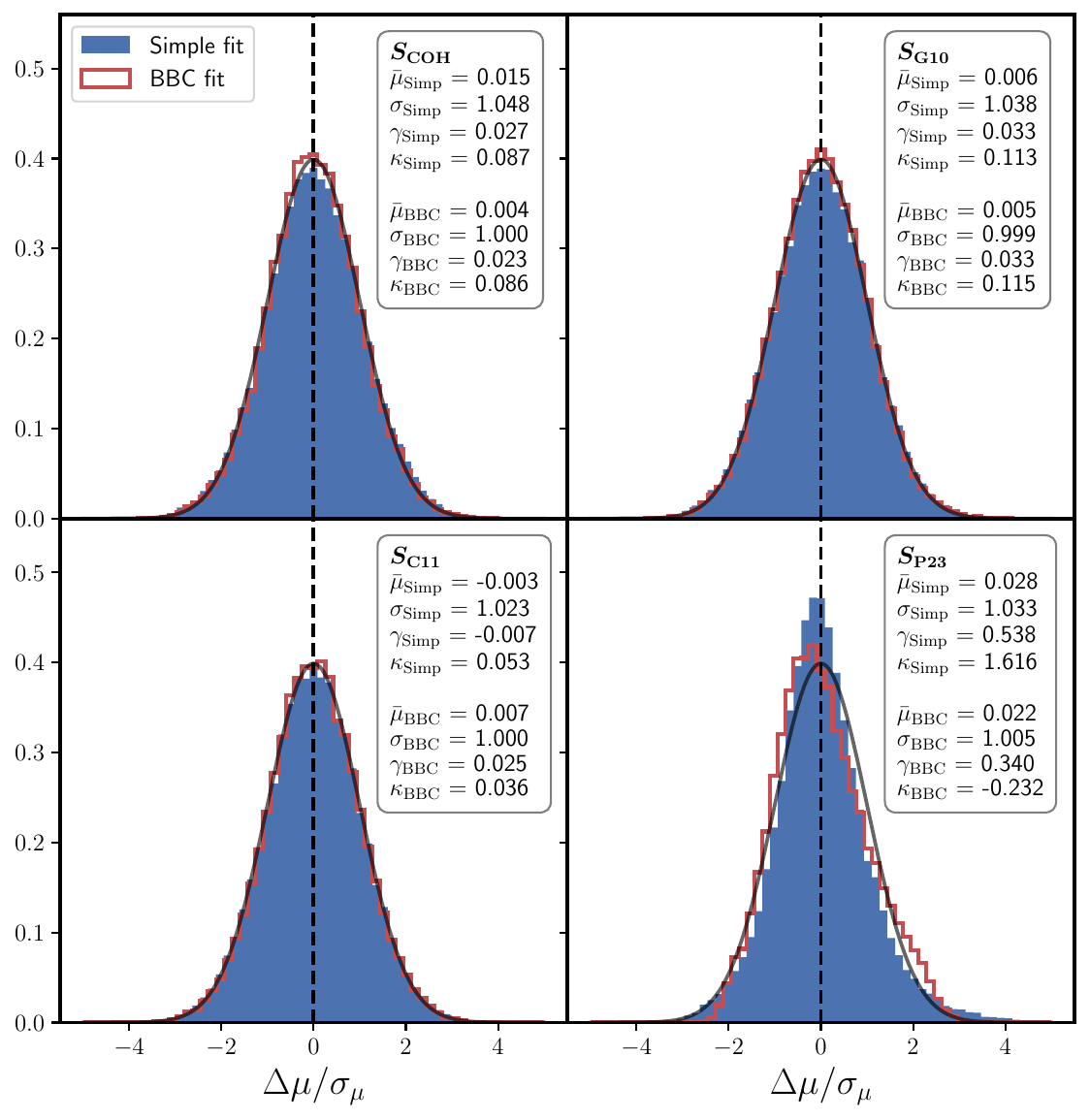} 
        \includegraphics[width=0.45\textwidth]{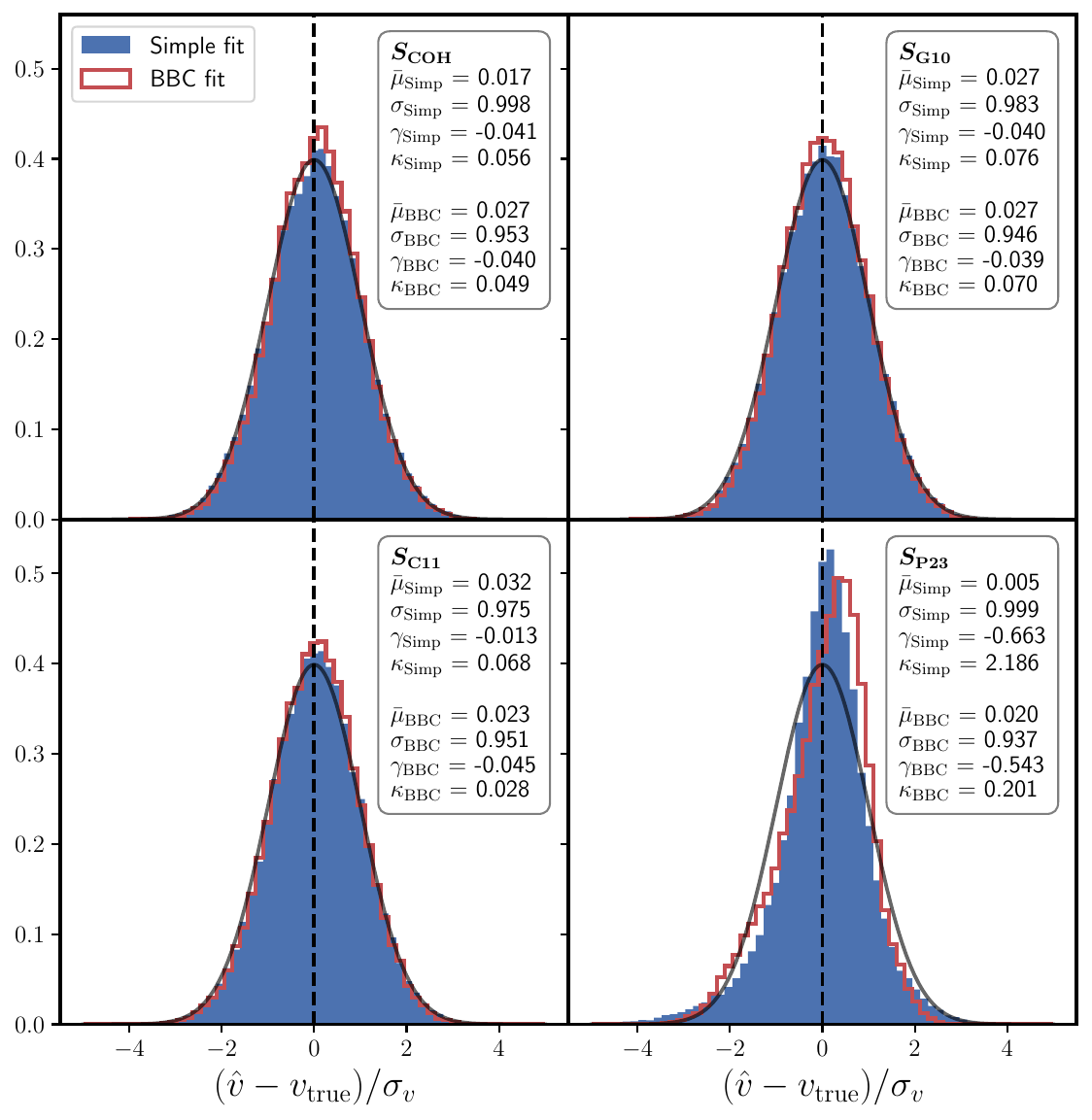} 
    \caption{Histograms of the pulls of the Hubble diagram residuals (left) and velocity residuals (right). Results for the simple fit are in blue and results for the BBC fit are in red. The standard deviation $\sigma$, the skewness $\gamma$ and the excess kurtosis $\kappa$ of the samples is given. A Gaussian function is plotted in black.}
    \label{fig:pulls}
\end{figure*}
This trend is further seen in the pulls of the residuals and velocities in Fig.~\ref{fig:pulls}. The pulls of Hubble residuals and velocities are similar for the scatter models of \simu{COH}, \simu{G10}, and \simu{C11} when using both of the two fitting methods. For the \simu{P23} simulation, visual inspection shows that the pull distribution deviates from a Gaussian. 
To quantify this, we compute the skewness $\gamma$ and excess kurtosis $\kappa$ of the distribution ($\gamma=\kappa=0$ for a Gaussian distribution). Although the skewness of the Hubble residuals distribution is close to zero for \simu{COH}, \simu{G10}, and \simu{C11}, for \simu{P23}
we measure $\gamma_\mathrm{Simp}\sim0.54$ when using the simple fitting method and $\gamma_\mathrm{BBC}=0.34$ when using the BBC fitting method. Similarly, the kurtosis values of the Hubble residuals distribution for \simu{COH}, \simu{G10}, and \simu{C11} are relatively low, ranging between $\sim$~0.04--0.12 compared to the value of $\sim1.62$ found for \simu{P23} when using the simple fitting method. This excess of kurtosis seems to be mitigated when using the BBC fitting method, with a value of $\sim-0.23$. The deviations from zero of the skewness and kurtosis excess in \simu{P23} point to the presence of non-Gaussianities. This is expected, since the \citetalias{brout_its_2021} intrinsic scatter model used in the \simu{P23} simulation has dust-based color dependence that introduces non-Gaussian scatter. 


\begin{figure}[h]
    \centering
    \includegraphics[width=\columnwidth]{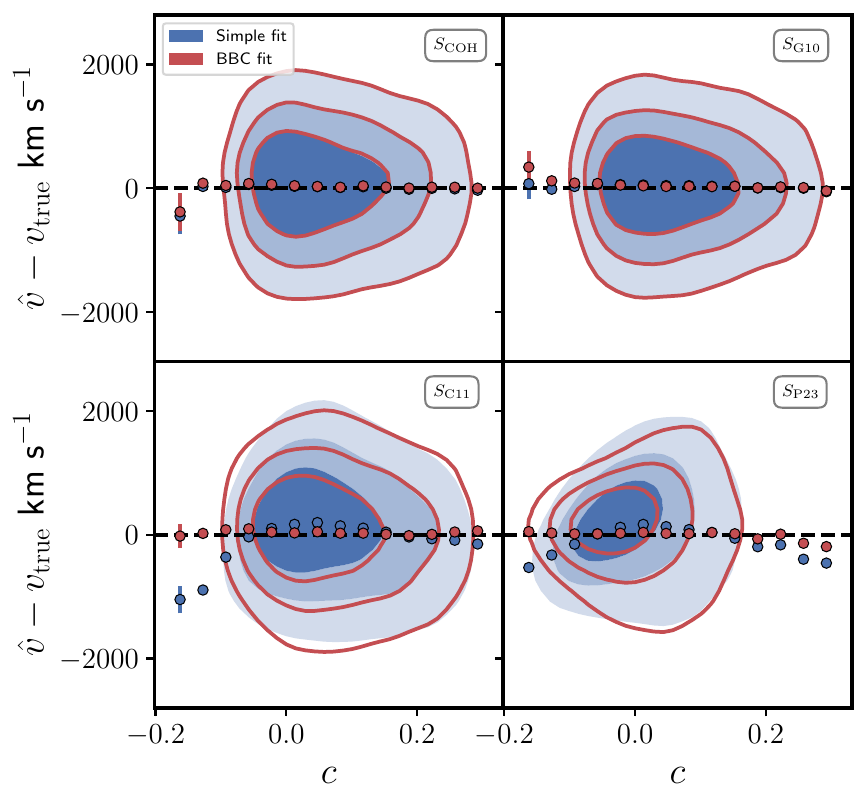}
    \caption{Bias in estimated velocity as a function of fitted SALT color parameter $c$. The contours represent the 25\%, 50\% and 75\% density levels of the SN Ia sample and points are the velocity bias in different $c$ bins. Blue color is used for the simple fitting method and red for the BBC methods.}
    \label{fig:vbias_c}
\end{figure}

In Fig.~\ref{fig:vbias_c} we present the velocity residuals as a function of the recovered SALT color parameter $c$. For the simulations \simu{COH} and \simu{G10}, the two fitting methods perform similarly. For the two more chromatic models of \simu{C11} and \simu{P23}, the simple fit method presents a bias when $c<0$. For the simulation \simu{C11}, this bias increases linearly below $c\simeq0$ and reaches a value of $\sim-480$~km~s$^{-1}$ at $c \sim -0.1$. For the simulation \simu{P23}, a similar bias of $\sim-190$~km~s$^{-1}$ is observed when $c \sim-0.1$. Furthermore, in \simu{P23}, we see a bias for high values of $c$ with an amplitude of $\sim-180$~km~s$^{-1}$ at $c\sim0.2$. The BBC method avoids this bias through the correction performed in color bins, which corrects the average residuals toward a value of zero. We note that in \simu{P23}, the density contours show a correlation between the PV bias and the $c$ parameter; for BBC, this correlation seems to be slightly mitigated. However, for both methods, the distribution of the bias on PVs is asymmetric as a consequence of the non-Gaussianity of the residuals. We further discuss the non-Gaussianity of the HD residuals in Appendix~\ref{app:nonGauss}.

\subsection{$\fsig$ fit results for the four scatter models}
\begin{figure}[h]
    \centering
    \includegraphics[width=\columnwidth]{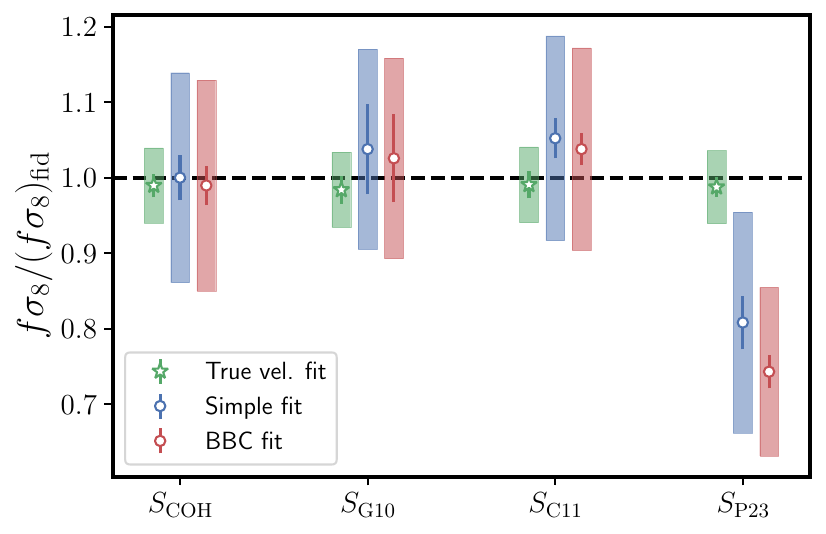}
    \caption{Average results over the eight mocks for our four different intrinsic scatter models. The points with error bars represent the mean $\fsig$ over the eight mocks, while the wider colored bands represent the mean error obtained over the 8 mocks. The results for true velocities are in green, the results of the simple fit are in blue, and the results of the BBC fit are in red.}
    \label{fig:resfs8}
\end{figure}
In Fig.~\ref{fig:resfs8} and Table~\ref{tab:resfs8}, we present the results of the fitting procedure for the four intrinsic scatter models. Fitting for $\fsig$ with the true velocities, which correspond to a test case, gives unbiased results for all models with a mean error of about $\sim5\%$ that represents the statistical limit of our simulated data. The fact that this fit is unbiased is a hint that the particular distribution of \sn\ host galaxy masses does not lead to a particular velocity statistic that could bias the $\fsig$ measurement. For the \simu{COH} simulation, as expected, the simple and the BBC fitting procedure perform largely the same with an unbiased value and an error of $\sim14\%$. For the \simu{G10} and \simu{C11} models, the two fitting procedures perform also very similarly with error on $\fsig$ of about $13\%$. However, for the \simu{C11} simulation, a bias of $\sim4\%$ with a significance of $\sim2\sigma$ seems to be present in both the simple and the BBC fit. This bias is negligible when compared to the $13\%$ averaged uncertainty. For our most realistic simulation, \simu{P23}, the results for the two fitting methods are biased with a strong significance. The simple method is slightly less biased, with a bias of $\sim-20\% $ against a bias of $\sim-26\%$ for the BBC fit. We understand this bias as a consequence of the non-Gaussianity of the estimated velocities noted in the previous section. For the mean error on $\fsig$ we obtain $\sim11\%$ from the BBC fit and $\sim15\%$ from the simple fit. 

Table~\ref{tab:resfs8} shows also that, for the \simu{COH}, \simu{C11} and \simu{P23}, the mean errors on $\fsig$ obtained from the fit seem to be slightly higher than the scattering of the measures computed over the eight mocks. This could point to an overestimation of the errors that should be investigated in future works. This effect does not appear for the \simu{G10} simulation, which shows the opposite behavior, with a mean error smaller than the scatter. We note that while the scattering of the results over our eight mocks is similar for the simple and the BBC fit in the \simu{COH} and \simu{G10} simulations, with values of $\sim7.5\%$ for \simu{COH} and $\sim16\%$ for \simu{G10}. In the case of the most chromatic models \simu{C11} and \simu{P23} the scatter is less important for the BBC fit.For \simu{C11} we find a scatter of the results of about $\sim7\%$ for the simple fit and of about $\sim6\%$ for the BBC fit, for \simu{P23} we find a scatter of $\sim12\%$ for the simple fit and $\sim9\%$ for the BBC fit. We note that these results are using the low statistics of only eight realizations and may also be subject to the small correlations existing between the sub-mocks.
\begin{table*}
    \caption{Results of the $\fsig$ fit for our four simulations. For each of the fits we show the average fitted value of $\fsig$, the average uncertainty on $\fsig$ and the standard deviation of $\fsig$ results across our eight mocks.}\label{tab:resfs8}
    \centering
    \renewcommand{\arraystretch}{1.5}
    \scriptsize
    \begin{tabular}{c||c|c|c||c|c|c||c|c|c||c}
\multirow{2}{*}{Models} & \multicolumn{3}{c||}{True fit}                                                                                        & \multicolumn{3}{c||}{Simple fit}                                                                                    & \multicolumn{3}{c||}{BBC fit}                                                                                         & \multicolumn{1}{c}{}  \\ 
\cline{2-10}
                        & $\langle\fsig\rangle / \left(\fsig\right)_\mathrm{fid}$ & $\sqrt{\langle\sigma_{\fsig}^2\rangle}$ & STD($\fsig$) & $\langle\fsig\rangle / \left(\fsig\right)_\mathrm{fid}$ & $\sqrt{\langle\sigma_{\fsig}^2\rangle}$ & STD($\fsig$) & $\langle\fsig\rangle / \left(\fsig\right)_\mathrm{fid}$ & $\sqrt{\langle\sigma_{\fsig}^2\rangle}$  & STD($\fsig$) & \multicolumn{1}{c}{}\\[1ex]
\hhline{=#=|=|=#=|=|=#=|=|=#}
\simu{COH} & 0.990 $\pm$ 0.016 & 5.0\% & 4.5\% & 1.000 $\pm$ 0.030 & 13.9\% & 8.4\% & 0.990 $\pm$ 0.026 & 14.0\% & 7.5\%\\
\simu{G10} & 0.984 $\pm$ 0.018 & 5.0\% & 5.3\% & 1.038 $\pm$ 0.059 & 13.3\% & 16.2\% & 1.026 $\pm$ 0.058 & 13.3\% & 16.0\%\\
\simu{C11} & 0.991 $\pm$ 0.018 & 5.0\% & 5.2\% & 1.052 $\pm$ 0.026 & 13.6\% & 7.1\% & 1.038 $\pm$ 0.021 & 13.4\% & 5.8\%\\
\simu{P23} & 0.988 $\pm$ 0.013 & 4.8\% & 3.7\% & 0.808 $\pm$ 0.036 & 14.6\% & 12.4\% & 0.743 $\pm$ 0.022 & 11.2\% & 8.4\%\\[1ex]
\hhline{----------}
\end{tabular}

\end{table*}

\subsection{Propagation of the uncertainty on the BS21 intrinsic scatter model parameters to $\fsig$}
Here, we test the impact of adding the covariance matrix that represents the uncertainty on the \citetalias{brout_its_2021} model parameters. We compute this covariance matrix in the same way as  stated in Sect.~\ref{sec:method:bbc} of \citetalias{vincenzi_dark_2024}.
During the construction of the \citetalias{brout_its_2021} intrinsic scatter model covariance matrix, an extra cut is added on the \sn\ sample by BBC. This cut requires that the corrected distance modulus passes the cuts presented in Sect.~\ref{sec:method:samplecuts} for each of the bias corrections obtained from the different bias correction simulations. This cut slightly reduces our sample from $\langle N_\mathrm{SN}\rangle \sim 6916$ to $\langle N_\mathrm{SN}\rangle \sim6630 $.

The fit results are presented in Table~\ref{tab:fs8sysres} as an average over the eight mocks. The fit of the true velocities gives, as before, unbiased results with a relative error of 4.9\%. The sub-sampling due to the additional cut slightly worsens the simple fit bias to a value of $\sim-29\%$ with the a slightly higher relative error of $\sim16\%$. The bias of the BBC fit without the covariance matrix is slightly increased by the sub-sampling  with a value of $\sim-35\%$ and a slightly higher relative error of $\sim 12\%$. The BBC fit that includes the intrinsic scatter model covariance matrix does not show any significant variations from the one without the inclusion of the covariance matrix, with an average shift of $\sim-0.2\%$ and no changes on the $\fsig$ error. From this, we conclude that the error on $\fsig$ is not dominated by the realizations of the P23 scatter model parameters, as when measuring $w$, but rather the amplitude of the Hubble residual scatter. This difference with respect to other cosmological analyses such as the constraints on the dark energy equation of state can be explained by the fact that in the $\fsig$ analysis the statistical error is scaled down by the density of \sns\ and not by their total number. Hence, the intrinsic scatter model systematic does not contribute significantly to the error budget for $\fsig$.
\begin{table*}
    \centering
    \caption{Results for the \simu{P23} simulation with and without including the intrinsic scatter covariance matrix from \citetalias{brout_its_2021} parameter uncertainty. The first column shows the average of $\fsig$ fitted values and the second column show the average uncertainty on $\fsig$ obtain over our eight mocks.}\label{tab:fs8sysres}
    \renewcommand{\arraystretch}{1.5}

    \begin{tabular}{c||c|c}
        & $\langle\fsig\rangle / \left(\fsig\right)_\mathrm{fid}$ & $\sqrt{\langle\sigma_{\fsig}^2\rangle} $ \\ 
        \hhline{=#==}
        True fit & 0.986 $\pm$ 0.017 & 4.9\%\\
        Simple fit & 0.713 $\pm$ 0.039 & 15.7\%\\
        BBC fit & 0.649 $\pm$ 0.028 & 12.2\%\\
        BBC + int. scat. cov fit & 0.647 $\pm$ 0.028 & 12.2\%\\[1ex]
        \hline
    \end{tabular}
\end{table*}

\subsection{The $\sigma_u$ systematic}\label{sec:res:sigu}
\begin{figure}[h]
    \centering
    \includegraphics[width=\columnwidth]{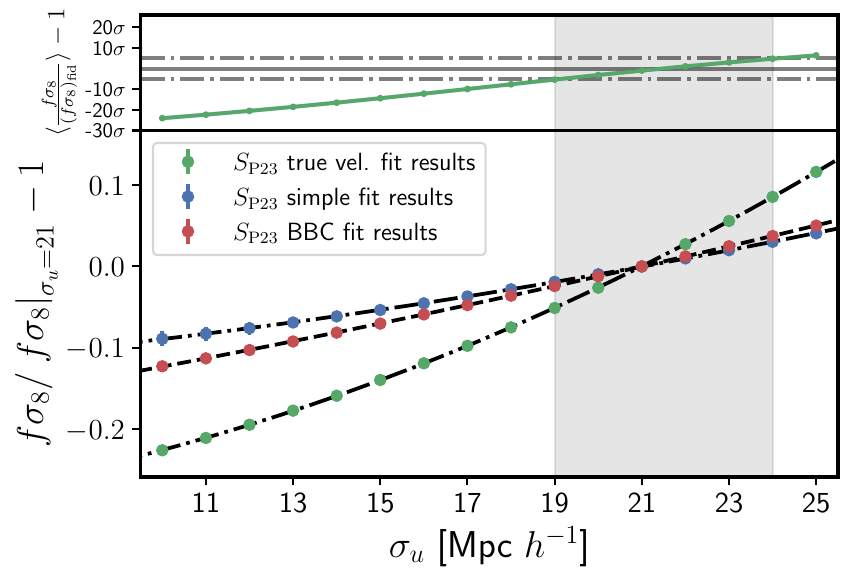}
    \caption{Results of the fit for different value of the damping parameter $\sigma_u$. True vel. fit is represented in green, the simple fit is in blue and the BBC fit is represented is in red. The top panel represent the pull for the true vel. fit in term of $\sigma$, the plain and dotted grey line represents respectively the 0 and 5$\sigma$ intervals. The bottom panel represents the relative variation of $\fsig$ with respect to the $\sigma_u=21$~Mpc~$h^{-1}$ case with the dashed black line corresponding to second degree polynomial fit. The grey color-span correspond to the value of $\sigma_u$ compatible with the fiducial $\fsig$ in a 5$\sigma$ interval with respect to the true vel. fit.}
    \label{fig:fs8_sigu}
\end{figure}
As stated in Sect.~\ref{sec:method}, we decide to fix the damping parameter to the value of $\sigma_u=21$~Mpc~$h^{-1}$. To test the robustness of our nominal choice for the fit of $\fsig$ we re-fit the \simu{P23} simulation with values of $\sigma_u$ ranging from 10~Mpc~$h^{-1}$ to 25~Mpc~$h^{-1}$. The results are presented in Fig.~\ref{fig:fs8_sigu}. We observe a linear correlation between $\fsig$ and the fixed value of $\sigma_u$. This correlation can be understood as a consequence of the degeneracy between $\fsig$ and $\sigma_u$: a higher $\sigma_u$ will give a stronger damping that will lower the value of the integral of Eq.~\ref{eq:cvvDu}, this lower value can be compensated for by a higher $\fsig$. We find this correlation to be stronger in the true velocity fit than for the BBC and simple fit. We adjust the relation between the variation of $\fsig$ and $\sigma_u$ by a second order polynomial. Then we can estimate the $\sigma_u$ systematic by 
\begin{equation}\label{eq:syssigu}
    \left(\sigma_{\fsig}^{\sigma_u}\right)^2 \simeq \left(\left. \frac{\partial\fsig}{\partial\sigma_u}\right|_{\sigma_u=21~\text{Mpc}~h^{-1}}\right)^2 (\Delta\sigma_u)^2.
\end{equation}
Using the average of the result of the $\fsig$ fit with the true velocity over the eight mocks, we find that the range of $\sigma_u$ that allows us to recover the fiducial value of $\fsig$ in a $5\sigma$ confidence range is $\sigma_u \in \left[19, 24\right]$~Mpc~$h^{-1}$. We note that  the best-fit $\sigma_u$ values found with the fit on true velocities is $\sigma_u \simeq 21.5$~Mpc~$h^{-1}$, slightly higher than the value we calibrated in Sect.~\ref{sec:method:pwsigu} and used as a baseline, $\sigma_u\simeq21$~Mpc~$h^{-1}$. The range we found is consistent with the one found in Table~1 of \cite{lai_using_2023} that is $\sigma_u \in \left[19, 23\right]$~Mpc~$h^{-1}$. By applying Eq.~\ref{eq:syssigu} on this range we find $\sigma_{\fsig}^{\sigma_u} \sim 6 \%$, contributing to roughly $\sim$23\% of the total error budget of $\fsig$. This result is again similar to the variation shown in Table~1 of \cite{lai_using_2023} in which $\fsig$ varies by $\sim7\%$ over a $\sigma_u$ range between 19~Mpc~$h^{-1}$ and 23~Mpc~$h^{-1}$.

    

\section{Conclusion}\label{sec:conc}
In this paper, we simulated eight realizations of the 10 years of \sn\ observations from the WFD program of the LSST survey for four different intrinsic scatter models. Our simulations include realistic distribution of \sn\ host masses and correlations between \sn\ SALT parameters and hosts. We build the \sn\ Hubble diagram and estimate velocities, comparing two different approaches. The simple method, used in \citetalias{carreres_growth-rate_2023}, is a \textit{naive} construction of the Hubble diagram from the Tripp relation. The second, the BBC approach, uses large simulations to correct for biases in the Hubble diagram. We show that the two methods have comparable estimates of PVs for the two less chromatic intrinsic scatter models. However, we find that for the two more chromatic models the simple approach slightly biases the velocities with respect to the color parameter, $c$. The BBC approach corrects this bias in PV estimates. 

We use the maximum likelihood method to fit for $\fsig$ and show that the fit with true velocities is not biased, indicating that the \sn\ host distribution does not lead to a biased measurement of $\fsig$. For three of our intrinsic scatter models, the achromatic, G10, and C11 models, we recover the fiducial value of $\fsig$ without bias and with an error on the order of $\sim14~\%$. In the case of the dust based model, both the simple and the BBC fitting methods present bias on $\fsig$ of about $\sim-20$\%. We interpret this as a consequence of the non-Gaussianities introduced by the dust term in the SN color distribution. This result motivates searches for new methods to take into account the non-Gaussian distribution of Hubble diagram residuals in the scope of an $\fsig$ measurement. That being said, this non-Gaussianity has still not been seen in current datasets and needs to be investigated, as it could become more significant with the increase of statistics and better calibration in the low redshift data of the new generation of surveys. 

Furthermore, we find that including the covariance due to the uncertainty on the \citetalias{brout_its_2021} model parameters does not significantly increase the error on $\fsig$, with a negligible contribution to the total error budget. Overall, the error budget of $\fsig$ is dominated by the statistical error that represents $>75\%$ of the total error on $\fsig$. We evaluate that the current systematic error is dominated by the uncertainty on the value of the nuisance parameter of the damping function $\sigma_u$. We quantify that the uncertainty on $\sigma_u$ results in a systematic error on $\fsig$ of about $\sim6$\%. This last result further motivates the ongoing effort to find a more physically motivated parametrization of the redshift space distortion effect on the velocity power spectrum \citep{dam_exploring_2021}.

\begin{acknowledgements}
This paper has undergone internal review in the LSST Dark Energy Science Collaboration. 
The internal reviewers were Dillon Brout and Tamara M. Davis.

The DESC acknowledges ongoing support from the Institut National de 
Physique Nucl\'eaire et de Physique des Particules in France; the 
Science \& Technology Facilities Council in the United Kingdom; and the
Department of Energy and the LSST Discovery Alliance
in the United States.  DESC uses resources of the IN2P3 
Computing Center (CC-IN2P3--Lyon/Villeurbanne - France) funded by the 
Centre National de la Recherche Scientifique; the National Energy 
Research Scientific Computing Center, a DOE Office of Science User 
Facility supported by the Office of Science of the U.S.\ Department of
Energy under Contract No.\ DE-AC02-05CH11231; STFC DiRAC HPC Facilities, 
funded by UK BEIS National E-infrastructure capital grants; and the UK 
particle physics grid, supported by the GridPP Collaboration.  This 
work was performed in part under DOE Contract DE-AC02-76SF00515.

J.E.B. acknowledges funding from Excellence Initiative of Aix-Marseille University - A*MIDEX, a French ``Investissements d'Avenir'' program (AMX-20-CE-02 - DARKUNI).
L.G. acknowledges financial support from AGAUR, CSIC, MCIN and AEI 10.13039/501100011033 under projects PID2023-151307NB-I00, PIE 20215AT016, CEX2020-001058-M, ILINK23001, COOPB2304, and 2021-SGR-01270.
\end{acknowledgements}

\begin{contribution}
B.~C. performed the simulations, the analysis, and wrote the majority of the paper. R.~C. contributed to the simulation setup, paper writing and review, and participated in analysis discussions. E.~R.~P. contributed to the writing and review of the paper and participated in analysis discussions. D.~S. contributed to the writing and review of the paper and participated in analysis discussions. C.~R. contributed to the code development and review. D.~R. reviewed the code and the manuscript. 
\end{contribution}

\software{\texttt{numpy} \citep{harris_array_2020}, \texttt{matplotlib} \citep{Hunter:2007},  \texttt{astropy} \citep{robitaille_astropy_2013, the_astropy_collaboration_astropy_2018, the_astropy_collaboration_astropy_2022}, \texttt{scipy} \citep{virtanen_scipy_2020}, \texttt{pandas} \citep{the_pandas_development_team_pandas-devpandas_2024}}

\bibliography{references}{}
\bibliographystyle{aasjournalv7}

\appendix

\section{Fit results for the SNe~Ia standardization parameters}\label{app:hdpar}
In Table~\ref{tab:stdpar} we show the results of the fit for the \sn\ standardization parameters. For the stretch correction linear coefficient $\alpha$, we find that for the \simu{COH}, \simu{G10} and \simu{C11}, both the simple fit and the BBC fit recover the input value of $\alpha=0.15$ within the mean uncertainty range. In the case of the \simu{P23} simulation, the BBC method results in a significant bias of $\sim-0.01$ with respect to the input $\alpha$ value, while the simple method is unbiased.

For the color correction linear coefficient $\beta$, the fitted values for the \simu{COH} and \simu{G10} simulations are slightly underestimated when using the simple method, while the values obtained with the BBC method are compatible with the input value of $\beta=3.1$ within the averaged uncertainty. In the case of the \simu{C11} simulation the simple method gives an underestimated value when compared to the input value of $\beta=3.8$, while the BBC method value is compatible with the input value within the averaged uncertainty. For the \simu{P23} simulation, there is no input value of $\beta$; however the value recovered by both the simple and the BBC method seem compatible and are similar to the value of $\beta\sim2.83$ found on DES-SN+low-z simulations in Table~8 of \citetalias{vincenzi_dark_2024}.

Finally, for the mass step parameter $\gamma$, we find that for the \simu{COH}, \simu{G20} and \simu{C11} simulations, the simple fit recovers the value of the input $\gamma=0.050$ within the average uncertainty error. The BBC fit recovers the value of $\gamma$ for the \simu{C11} simulation but  presents a bias of $\sim1.5\sigma$ with respect to the averaged uncertainty for the \simu{COH} and \simu{G10} simulations. In the case of the \simu{P23} there is no input value for $\gamma$ since the mass-step emerges from the dust extinction-host mass correlation. The value of $\gamma$ fitted using the BBC method is close to a value of $\gamma\sim0$, a result similar to what was found in DES simulations in Table~4 of \citetalias{vincenzi_dark_2024}.

\begin{table}[h]
\centering
\caption{Results of the $\alpha$, $\beta$ and $\gamma$ fits for our four simulations. For each fit we show the average fitted value and the average uncertainty across our eight mocks.}
\label{tab:stdpar}
    \begin{tabular}{c||c|c||c|c|}
    \multirow{2}{*}{Models} & \multicolumn{2}{c||}{Simple fit}                                                                  & \multicolumn{2}{c|}{BBC fit}                                                                      \\ 
    \cline{2-5}
                            & $\langle\alpha\rangle$ & $\sqrt{\langle\sigma_{\alpha}^2\rangle}$ & $\langle\alpha\rangle$ & $\sqrt{\langle\sigma_{\alpha}^2\rangle}$  \\ 
    \hhline{=#==#==|}
    \simu{COH} & 0.1490 $\pm$ 0.0005 & 0.0013 & 0.1495 $\pm$ 0.0005 & 0.0013\\ 
    \simu{G10} & 0.1487 $\pm$ 0.0004 & 0.0013 & 0.1490 $\pm$ 0.0004 & 0.0012\\ 
    \simu{C11} & 0.1491 $\pm$ 0.0004 & 0.0014 & 0.1497 $\pm$ 0.0003 & 0.0013\\ 
    \simu{P23} & 0.1500 $\pm$ 0.0001 & 0.0012 & 0.1399 $\pm$ 0.0001 & 0.0008\\ 
    \hline
    \end{tabular}

    \vspace{3ex}
    
    \begin{tabular}{c||c|c||c|c|}
    \multirow{2}{*}{Models} & \multicolumn{2}{c||}{Simple fit}                                                                  & \multicolumn{2}{c|}{BBC fit}                                                                      \\ 
    \cline{2-5}
                            & $\langle\beta\rangle$ & $\sqrt{\langle\sigma_{\beta}^2\rangle}$ & $\langle\beta\rangle$ & $\sqrt{\langle\sigma_{\beta}^2\rangle}$  \\ 
    \hhline{=#==#==|}
\simu{COH} & 3.084 $\pm$ 0.005 & 0.015 & 3.098 $\pm$ 0.005 & 0.014\\ 
\simu{G10} & 3.064 $\pm$ 0.005 & 0.014 & 3.091 $\pm$ 0.005 & 0.013\\ 
\simu{C11} & 3.504 $\pm$ 0.005 & 0.016 & 3.700 $\pm$ 0.006 & 0.016\\ 
\simu{P23} & 2.875 $\pm$ 0.006 & 0.014 & 2.828 $\pm$ 0.003 & 0.011\\ 

    \hline
    \end{tabular}

        \vspace{3ex}
    
    \begin{tabular}{c||c|c||c|c|}
    \multirow{2}{*}{Models} & \multicolumn{2}{c||}{Simple fit}                                                                  & \multicolumn{2}{c|}{BBC fit}                                                                      \\ 
    \cline{2-5}
                            & $\langle\gamma\rangle$ & $\sqrt{\langle\sigma_{\gamma}^2\rangle}$ & $\langle\gamma\rangle$ & $\sqrt{\langle\sigma_{\gamma}^2\rangle}$  \\ 
    \hhline{=#==#==|}
    \simu{COH} & 0.049 $\pm$ 0.001 & 0.003 & 0.046 $\pm$ 0.001 & 0.003\\ 
    \simu{G10} & 0.048 $\pm$ 0.001 & 0.003 & 0.045 $\pm$ 0.001 & 0.003\\ 
    \simu{C11} & 0.054 $\pm$ 0.001 & 0.003 & 0.048 $\pm$ 0.001 & 0.003\\ 
    \simu{P23} & 0.079 $\pm$ 0.002 & 0.003 & -0.004 $\pm$ 0.001 & 0.002\\
    
    \hline
    \end{tabular}
\end{table}

\section{Non-Gaussianties and variations of the P23 parameters}\label{app:nonGauss}
\begin{figure}
    \centering
    \includegraphics[width=\textwidth]{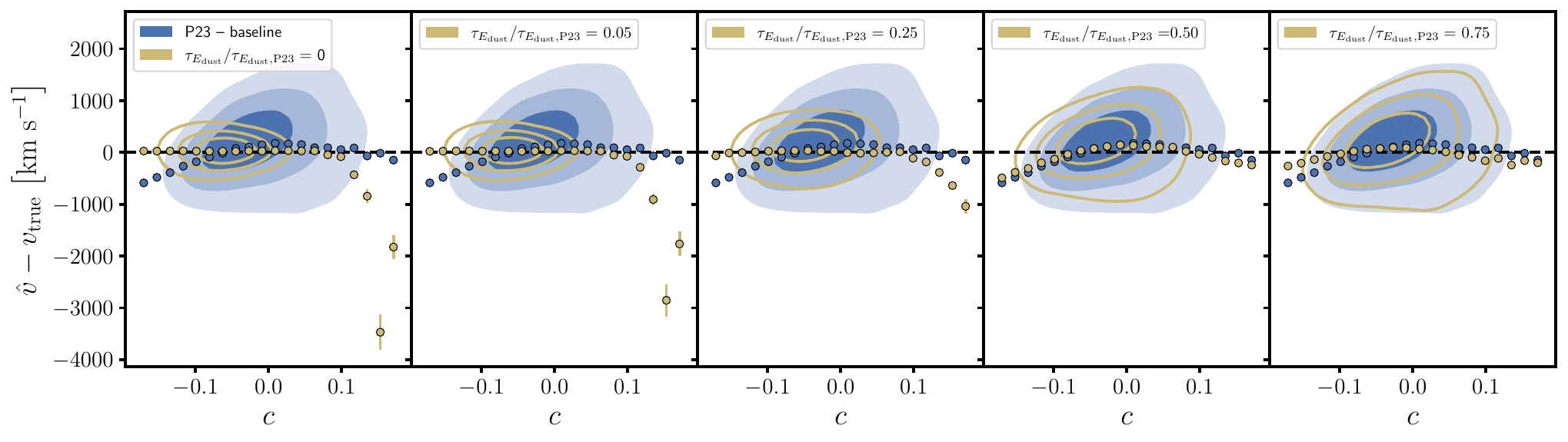}
    \caption{Bias in estimated velocity as a function of fitted SALT color parameter $c$. The contours represent the 25\%, 50\% and 75\% density levels of the SN Ia sample and points are the velocity bias in different $c$ bins. The blue contours correspond to \simu{P23}, while yellow contours are obtained using variation of the $\tau_\mathrm{E_{dust}}$ parameter.} \label{fig:vb_tau}
    \label{fig:placeholder}
\end{figure}

To further understand the origin of non-Gaussianities arising from the P23 dust model, we investigate the consequences of varying some parameters of the model. In particular, we simulate our sample with a reduced variance $\tau_{E_\mathrm{dust}}$ of the dust-extinction exponential distribution. We simulated the SNe~Ia sample for different values of $\tau_{E_\mathrm{dust}}$ reduced by 25, 50, 75, 95, and 100\% with respect to the value used in our nominal \simu{P23} simulation. In Fig.~\ref{fig:vb_tau} we show the color -- velocity-bias density contours for these different simulations, compared to that of \simu{P23}. We can see that the color bias and the asymmetry of the velocity distribution are growing along with the increase of the $\tau_\mathrm{E_\mathrm{dust}}$ parameter. 

\begin{figure}
    \centering
    \includegraphics[width=0.5\linewidth]{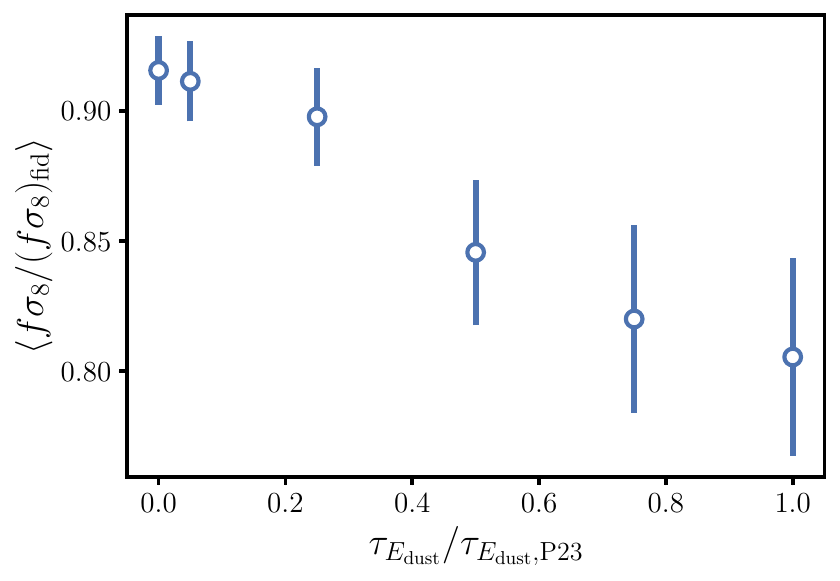}
    \caption{Average result over the eight mocks as a function of variations of the value of the $\tau_{E_\mathrm{dust}}$ with respect to the baseline value used in \simu{P23}.}
    \label{fig:fs8_tau}
\end{figure}

We fitted for $\fsig$ with these simulations using the simple fitting method. We did not run the BBC fitting method since it would require simulating a bias correction for each variation of $\tau_{E_\mathrm{dust}}$. For each mock, simulations of variation of $\tau_{E_\mathrm{dust}}$ use the same random seed. Thus, the host galaxies and velocities used in the fit of $\fsig$ are the same and the differences lie in the fitted SALT parameters. The results for $\fsig/(\fsig)_\mathrm{fid}$ are shown in Fig.~\ref{fig:fs8_tau} as a function of the ratio between the simulated $\tau_{E_\mathrm{dust}}$ and our baseline value used in \simu{P23}. We see that the increase of $\tau_{E_\mathrm{dust}}$ is correlated with a decrease of the bias in the fitted value of $\fsig$. In the ``No dust'' case ($\tau_{E_\mathrm{dust}}=0$) we remark that we still have a bias of $\sim -9\%$ with respect to the fiducial $\fsig$ value. Since the Hubble scatter is decreasing with the suppression of dust, this remaining bias could be due to other systematic effects that dominate this unrealistic low-scatter regime, and we let there identifications to later studies. We also tested reducing the scatter $\sigma_\beta$ of the distribution of $\beta$ but we found its impact on the $\fsig$ bias to be negligible.
\end{document}